\documentclass[11pt,oneside,letterpaper]{article}
\usepackage{amssymb}
\usepackage{amsmath}
\usepackage[dvips]{graphicx}
\usepackage{setspace}
\usepackage{amsfonts}
\usepackage{fancyhdr}
\usepackage{xcolor}
\usepackage{graphicx}
\usepackage{rotating}
\usepackage{comment}
\usepackage{color}
\usepackage{cite}
\usepackage{braket}
\usepackage{caption}
\usepackage{wrapfig}
\usepackage[utf8]{inputenc}
\usepackage[c]{esvect}
\usepackage{bbold}
\usepackage{graphicx,caption}

\definecolor{darkgreen}{rgb}{0,0.5,0}
\definecolor{darkblue}{rgb}{0,0,0.6}
\definecolor{purple}{rgb}{0.4,.2,0.7}

\renewcommand{\ket}{\rangle}
\renewcommand{\bra}{\langle}
\newcommand{\be}{\begin{equation}}
	\newcommand{\ee}{\end{equation}}

\newcommand{\B}{{\scaleto{B}{4.5pt}}}
\newcommand{\A}{{\scaleto{A}{4.5pt}}}

\newcommand{\bz}{{\boldsymbol z}}
\newcommand{\by}{{\boldsymbol y}}
\newcommand{\bx}{{\boldsymbol x}}
\newcommand{\bv}{{\boldsymbol v}}
\newcommand{\bn}{{\boldsymbol n}}

\usepackage[colorlinks=true,citecolor=darkgreen,linkcolor=black,urlcolor=purple]{hyperref}

\usepackage{pdfsync}
\makeatletter
\newcommand*{\defeq}{\mathrel{\rlap{%
			\raisebox{0.3ex}{$\m@th\cdot$}}%
		\raisebox{-0.3ex}{$\m@th\cdot$}}%
	=} 
\makeatother

\def\be{\begin{eqnarray}}
	\def\ee{\end{eqnarray}}

\newcommand{\bea}{\begin{eqnarray}}
	\newcommand{\eea}{\end{eqnarray}}
\def\ben{\begin{equation}}
	\def\een{\end{equation}}

     \let\r=v

\def\be{\begin{equation}}
	\def\ee{\end{equation}}
\def\ba{\begin{array}}
	\def\ea{\end{array}}

\def\ba#1\ea{\begin{align}#1\end{align}}
\def\bs#1\es{\begin{split}#1\end{split}}

\usepackage{scalerel}

\allowdisplaybreaks  % allow page breaks in displayed eqs

\numberwithin{equation}{section}
\def \be {\begin{equation}}
	\def \ee {\end{equation}}

\begin{document}
	\onehalfspacing
	
	\begin{center}

~
\vskip5mm

{\LARGE  
Relativity of the event: \\ examples in JT gravity and linearized GR\\
\ \\
}

Francesco Nitti$^1$, Federico Piazza$^2$ and  Alexander Taskov$^2$

\vskip5mm
{\it $^1$Universit\'e Paris Cit\'e, CNRS, Astroparticule et Cosmologie, F-75013 Paris, France. \\
$^2$Aix Marseille Univ, Universit\'{e} de Toulon, CNRS, CPT, Marseille, France. }

\vskip5mm

%{\tt piazza@cpt.univ-mrs.fr, a.tolley@imperial.ac.uk}

\end{center}

%	\begin{center}
%		
%		~
%		\vskip5mm
%		
%		
%		
%		{\LARGE  {Relativity 2.0 -- the Relativity of the Event.
%				\ \\
%		}}
%		
%		Federico Piazza$^1$ and  Alexander Taskov$^2$
%		
%		
%		\vskip5mm
%		%{\it Aix Marseille Univ, Universit\'{e} de Toulon, CNRS, CPT, Marseille, France
%			%} 
%		
%		\vskip5mm
%		
%		
%		%{\tt piazza@cpt.univ-mrs.fr }
%		
%		
%	\end{center}
	
	\vspace{4mm}
	
	\begin{abstract}

Observables in quantum gravity are famously defined asymptotically, at the boundary of AdS or Minkowski spaces. However, by gauge fixing a coordinate system or suitably dressing the field operators, an approximate,  ``quasi-local" approach is also possible,  that can give account of the measurements performed by a set of observers living inside the spacetime. 
In particular, one can attach  spatial coordinates to the worldlines of these observers 
and use their proper times as a time coordinate. 	Here we highlight that any such local formulation has to face the \emph{relativity of the event}, in that  
changing frame (= set of observers)  implies a reshuffling of the point-events and the way they are identified. As a consequence, coordinate transformations between different frames become \emph{probabilistic} in quantum gravity. 
We give a concrete realization of this mechanism in Jackiw-Teitelboim gravity, where a point in the bulk can be defined operationally with geodesics anchored to the boundary. We describe different ways to do so, each corresponding to a different \emph{frame}, and compute the variances of the transformations relating some of these frames. In particular, we compute the variance of the location of the black hole horizon, which appears smeared in most frames.  We then suggest how to calculate this effect in Einstein gravity, assuming knowledge of the wavefunction of the metric. The idea is to expand the latter on a basis of semiclassical states. Each element of this basis enjoys standard/deterministic coordinate transformations and the result is thus obtained by superposition.  As a \emph{divertissement}, we sabotage Lorentz boosts  by adding to Minkoswki space a quantum superposition of gravitational waves and compute the probabilistic coordinate transformation to a boosted frame at linear order. Finally, we attempt to translate the relativity of the event into the language of dressed operators.

	 \end{abstract}
%\vspace{.2in}
%\vspace{.3in}

\pagebreak
\pagestyle{plain}

\setcounter{tocdepth}{2}
{}
\vfill

\ \vspace{-2cm}
\newcommand{\deq}{{\overrightarrow {\Delta x}}^{\, 2}}
\renewcommand{\baselinestretch}{1}\small
\tableofcontents
\renewcommand{\baselinestretch}{1.15}\normalsize

\section{Introduction}

When first approaching special relativity one is forced to adapt their physical intuition to the absence of an absolute time. One way to do so is to focus on the individual event. In Einstein's \emph{gedanken} constructions all observers at least agree on what an event is---they just assign different space\emph{time} coordinates to it.   
  Events remain central building blocks in classical general relativity and quantum field theory, where they become labels for the local field operators. In this paper we expand on the fact that, in the presence of dynamical gravity, events cannot survive as absolute, observer-independent concepts. That is, different frames must be related by \emph{probabilistic} coordinate transformations.

Roughly speaking, the point is that the wave-function of the universe associates a probability amplitude to a statistical ensemble of off-shell spacetime  geometries and, because of gauge invariance,  there is no unique prescription for identifying \emph{which event corresponds to which} among these geometries. Different prescriptions lead to different definitions of event. One could rephrase the 
above in terms of local field operators. To make the latter gauge invariant one has to \emph{dress} them with some non-local combinations of the metric field~\cite{Donnelly:2015hta,Donnelly:2016rvo,Giddings:2018umg,Giddings:2019wmj}.
This metric dependence implies that the events which label those operators acquire the uncertainty of the fluctuating metric.

\subsection{Physical observers and local viewpoint} \label{sec:intro1}
In order to stay away from the above complications one could resort to studying only scattering amplitudes in asymptotically Minkowski space, or boundary correlators in asymptotically AdS. These are arguably the best defined observables in quantum gravity.  However, as cosmological observers, we do not enjoy the asymptotics  of Minkowski or AdS, and still we seem able to perform measurements and observations of the outside world. By modeling a set of free-falling observers living inside a spacetime, it must be possible to predict the outcomes of such measurements,  even in the presence of sizable metric fluctuations. 
In some approximation observers can be identified as worldlines, 
\begin{equation} \label{observers}
S_{\rm obs} \ = \ \sum_i \int dt_i \left[- m_i + g \sigma_i  \phi(X_i^\mu(t_i)) + j_i(t_i)\phi(X_i^\mu(t_i))\right]\, .
\end{equation}
The first term in the square bracket determines the approximately geodesic trajectory of the $i^{th}$ observer, $X_i^\mu(t_i)$,  with $t_i$ their proper time. The remaining two terms describe experiments. The first is the interaction Lagrangian of an Unruh-DeWitt detector of small coupling $g$ attached to each observer. The magnetic monopole $\sigma_i$ allows the detector to jump into an excited internal state when detecting a particle of the light field $\phi$. The second equips the observers with a classical source $j_i(t_i)$  generating $\phi$ particles at around some prearranged value (say, $t_i^{emit}$) of their proper time.

\begin{figure}[h]
%\vspace{-1cm}
\begin{center}
     \includegraphics[width=11cm]{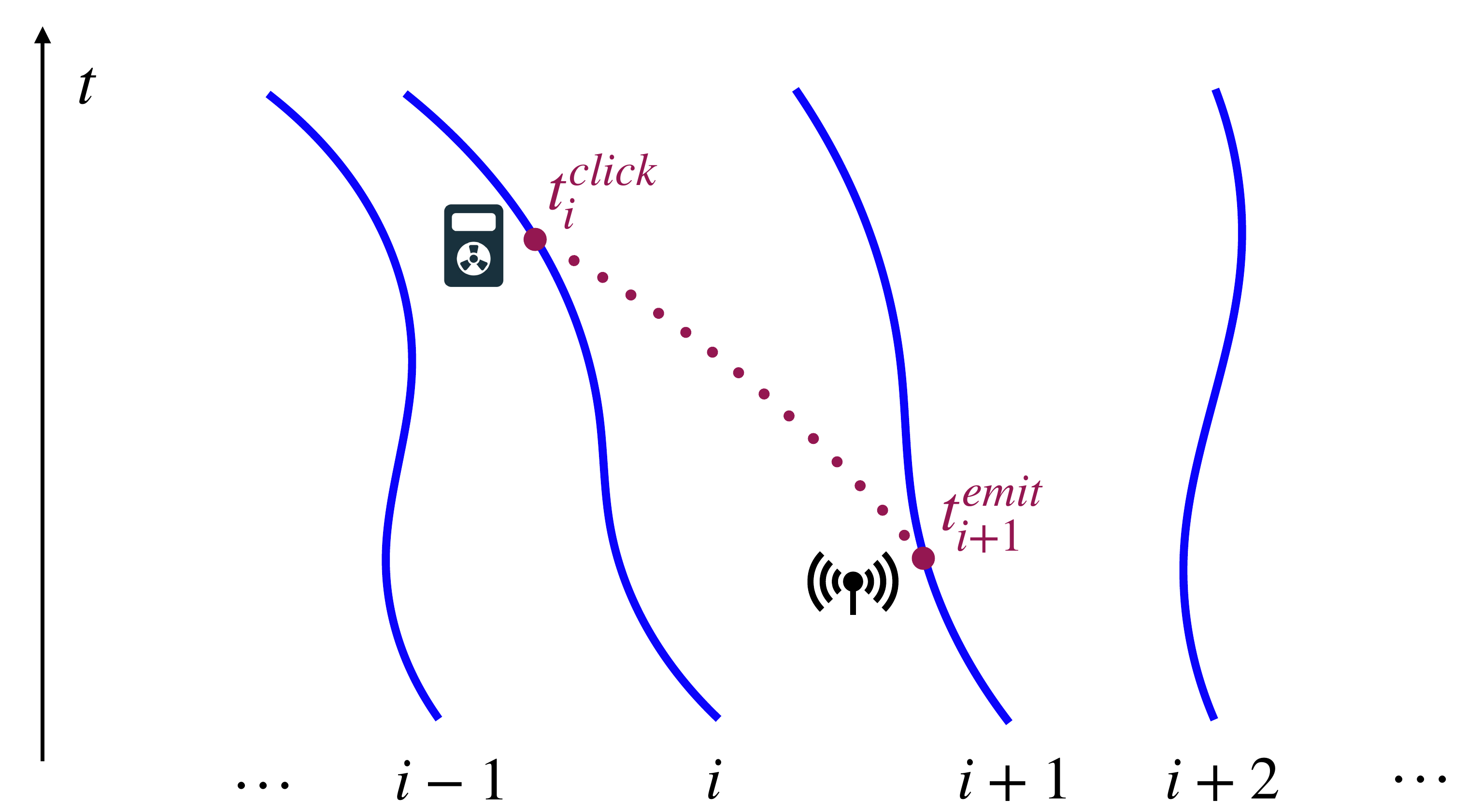}
  \end{center}\vspace{-.4cm}
  \captionsetup{width=.95\linewidth}
\caption{\small The type of gedanken experiments one can aim to describe with the observers model~\eqref{observers} coupled to dynamical gravity. The observer $i+1$ produces some light quanta at $t_{i+1}^{emit}$ with the aid of a classical source. One of these particles can be collected by the detector of the $i^{th}$ observer at some value of their proper time $t_i^{click}$. } \label{fig-1}
  \end{figure}

 There is no obstruction from considering the system gravity + observers in situations that deviate from classicality.\footnote{If anything, some low-energy restriction should be respected:  curvature should not grow so large as to destroy the observers and the metric should remain the only relevant gravitational degree of freedom.} For example, one could think of a first order cosmological phase transition in which the gravitational field is sourced by a scalar which tunnels through a potential barrier. We do not expect the wavefunction of the metric to remain peaked around a classical configuration during this process. 
 However, even in the absence of a classical spacetime, we can just \emph{define} an event as the clicking of the $i^{th}$ observer's detector and calculate the proper time $t_i^{click}$ at which this transition happens, say, with 90\% probability. 
 Such an occurrence is defined intrinsically, by the density matrix of the subsystem-detector reaching some state. At the same time, this occurrence is also a \emph{point} in the frame of our observers, with definite coordinates $t_i^{click}$ and $i$. These \emph{are} effectively coordinates, which can be made continuous by populating the spacetime with a \emph{diluted fluid of observers}, as explained at length in~\cite{Piazza:2021ojr,Piazza:2022amf}. The discrete index $i$ is then replaced by the continuous variable $\vec x$ which together with the proper time $x^0$ form lower case coordinates $x^\mu$ adapted to this set of observers.
 
 The spatial coordinates $\vec x$ correspond to a \emph{unitary gauge} choice for the observers' fluid model~\cite{Dubovsky:2005xd}. By construction, the observers' positions are exact in this gauge. 
Of course, a complete coordinate fixing can be made in many other ways, for example by asking that the metric assumes a specific form. However, in any other frame ${x'}^\mu $, our observers will find themselves in a superposition of trajectories because of the metric fluctuations. As a consequence, the event of the detector clicking, which is simply a point $x^\mu_{click}$ in the frame of our observers, corresponds, in a different frame, to some coordinate ${x'}^\mu$ with some \emph{calculable probability}. We denote this probability distribution  as
\begin{equation} \label{probability}
p({x'}^\mu | x^\mu_{click}  )\, .
\end{equation}
 In other words, what is a sharp, point-like event in one frame appears \emph{smeared} in  other frames.

\begin{figure}[t]
\vspace{-1cm}
\begin{center}
        \includegraphics[width=15cm]{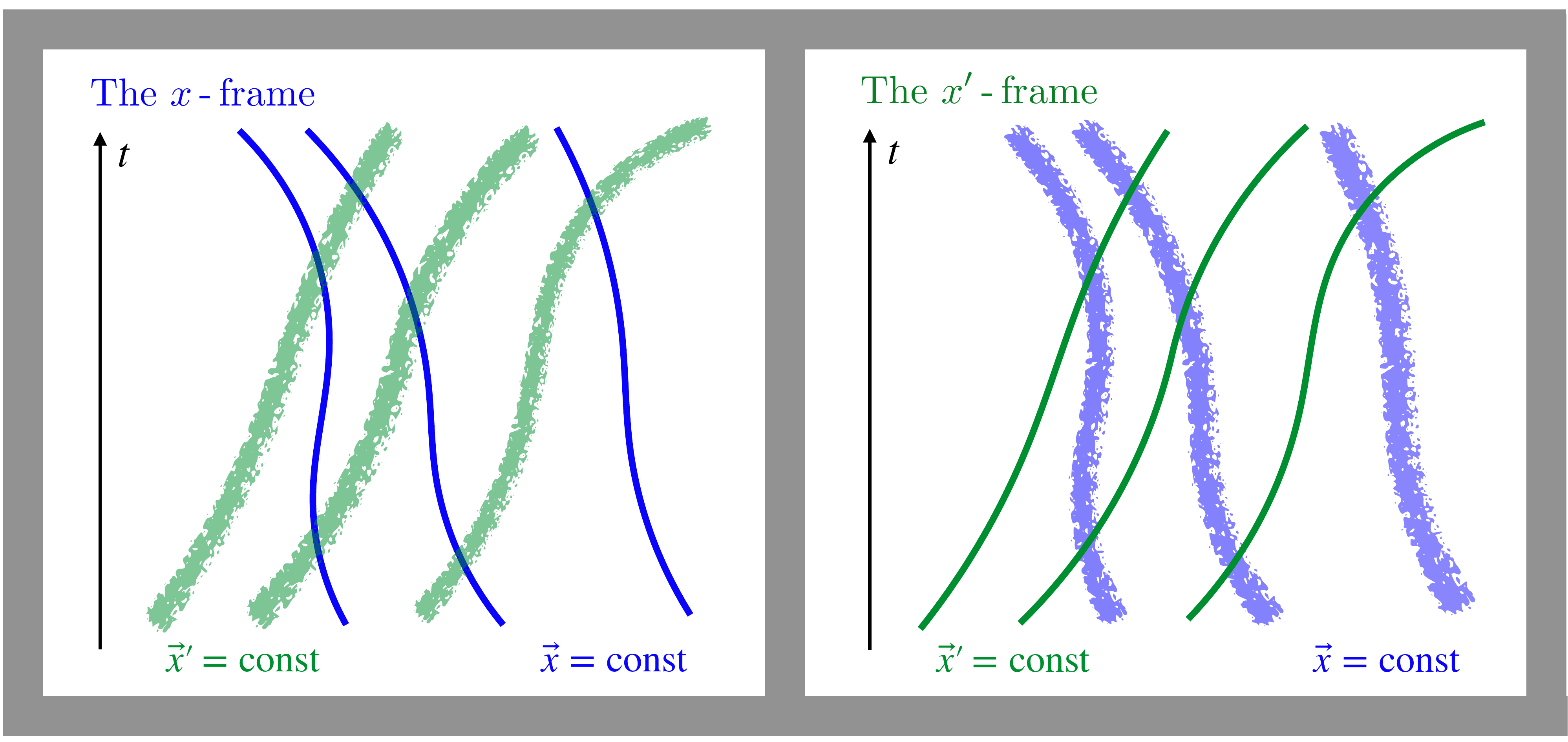}
  \end{center}%\vspace{-.4cm}
  \captionsetup{width=.95\linewidth}
\caption{\small Two sets of geodesic observers, blue and green, initially boosted with respect to each other, are used to define two different frames. The blue observers define the $x$-frame (left panel). By gauge choice, the blue worldlines are perfectly localized  in this frame, while the positions of the green observers are uncertain because of metric fluctuations. The same reasoning applies in reverse. The trajectories of the blue observers are uncertain in the $x'$-frame of the green observers (right panel). As a result, the transformations relating the $x$- and $x'$-frames are probabilistic.} \label{fig-frames}
  \end{figure}
The same reasoning applies, in particular, when the two frames represent two different sets of observers initially boosted one with respect to the other (Fig.~\ref{fig-frames}). The situation can be made even closer to the famous gendanken experiment of special relativity by assuming an initially Minkowski spacetime traversed by a quantum superposition of gravitational waves. In Sec.~\ref{boost}, we calculate the probabilistic coordinate transformation  relating the two boosted frames in this case.  Well beyond the relativity of simultaneity, any attempt at a local description of quantum gravity must face the \emph{relativity of the event}!

A comment on clock synchronization is perhaps in order at this point. The observers dynamics is invariant under $t_i \rightarrow t_i + \lambda_i$ which corresponds to embedding the same geodesic in the spacetime with a shifted value of the proper time. So, strictly speaking, while our observers' model~\eqref{observers} does define spatial coordinates, time is defined only up to a (space dependent) shift. For this reason, in~\cite{Chandrasekaran:2022cip,Witten:2023xze}, proper time is introduced as an independent dynamical variable. 
Alternatively, some kind of clock synchronization is needed to completely define a coordinate system.\footnote{We thank E. Witten for clarifications around these issues.} In this work we adopt a more mundane strategy: in Section \ref{JT}, where the setting is  asymptotically AdS spacetimes,  syncronization is imposed on the asymptotic (regulated) timelike boundary; in  Section \ref{prob}, which deals with  asymptotically flat spacetime,  we assume that the fluctuations of the metric vanish early on, giving us a place to safelyf anchor our geodesics and set the zeros of their proper times.

\subsection{Summary and results}

In this paper we explore the concepts outlined above in two different contexts, where these ideas can be made concrete and lead to more precise statements: 
\begin{enumerate}
\item 2-dimensional asymptotically AdS gravity (more precisely, the holographic Jackiw-Teitelboim (JT) model) 
\item  General Relativity in its linearized approximation around 4-dimensional Minkowski spacetime.
\end{enumerate}
Below we will briefly summarize our approach in these two situations.

In asymptotically AdS spaces, one can define a (gauge-invariant) anchoring procedure due to the existence of a boundary (and an associated dual field theory), from which one can send geodesics into the bulk to define a frame.   We  study such frames in 2-dimensional JT gravity  \cite{Almheiri2015-uf,Maldacena:2016hyu,Engelsoy:2016xyb, Maldacena:2016upp, Jensen:2016pah}, in which the geometry is locally $AdS_2$ and the only degree of freedom is a time-reparametrization function living on the boundary. The latter can be thought as the  degree of freedom of the $(0+1)$-dimensional dual  quantum mechanics, governed by a Schwarzian path integral.

As we will show,  if we define physical frames by  sending geodesics in from the boundary, the quantum uncertainty in the boundary position where the geodesics are anchored results in a corresponding uncertainty in the bulk coordinate system.

Concretely, we study quantum states which are fluctuations around classical 2d black hole geometries.  We will find that, when  quantizing the  boundary degree of freedom corresponding to the fluctuation,   the location of the horizon of a black hole looks smeared in most frames, with uncertainties that we compute explicitly.

More generally, we will see how  the coordinates of bulk events take on a probabilistic nature, with which one can associate and compute an uncertainty. In the Gaussian approximation for the dynamics of boundary fluctuations, this is equivalent to computing  the probabilistic coordinate transformation (as advocated in the previous section) between a boundary-anchored frame and any local  ``bulk''  frame (e.g. Poincar\'e coordinates of $AdS_2$).

We will use different prescriptions to define boundary-anchored bulk frames: 
\begin{enumerate}
\item A frame which uses null geodesics to define bulk coordinates in terms of boundary times. It was first introduced in \cite{Blommaert:2019hjr}, and we will refer to it as the {\em BMV frame}. As we will show,  this frame has the particular feature that the uncertainty in local coordinates is such that the causal structure of the spacetime remains deterministic (light-cones are not smeared).
\item A family of physical frames defined in terms of a collection of infalling observer\footnote{As a technical remark, these frames will depend on  a renormalization scale, which must be introduced to remove the boundary regulator. Different renormalization scale choices correspond to different definitions of the frame.}. The bulk coordinates defined in these frames are the boundary  time at which each observer leaves the boundary, and the proper time along the corresponding timelike geodesics\footnote{Such geodesics are not subject to metric fluctuation uncertainties, because the bulk geometry is locally AdS$_2$: however the coordinates defined this way are subject to the quantum fluctuations of the  boundary degree of freedom.} Unlike the BMV frame, here the causal structure of the spacetime becomes probabilistic, in the sense that any two events will have a non-trivial probability distribution of being connected by a light ray (as opposed to classical GR, in which light cones are sharp, and any  two given events are either null-separated, or they aren't, with probability one). This applies for example to the null curve describing the black hole horizon: we will compute in this frame the uncertainty in the proper time at which each observer crosses the horizon. 
\end{enumerate}

Our second  example of a system where we can compute  probabilistic coordinate transformations is 4-dimensional General Relativity in the linearized approximation around Minkowski spacetime. 
Unlike  JT gravity, where all the dynamics takes place on the boundary (and  quantum dynamics is computable in the boundary theory), in four dimensional asymptotically flat  Einstein gravity things are more complicated,  because of the absence of a holographic dual theory and of a timelike surface where one can anchor geodesics.

However, one can still construct a broad class  of gravitational states, at least in perturbation theory around Minkowski, and try to evaluate the corresponding probabilistic frame transformation between two sets of observers.  We do this by considering  a pure  quantum state in a superposition of coherent gravitational waves. In such a state, we will be able to construct the probabilistic coordinate transformation between two families of observers related to each other by an initial boost (in the vacuum state, i.e. in the absence of gravitational excitations, this is just a sharp Lorentz transformation). As we show explicitly, a spacetime event in one frame becomes, in the boosted  frame, a smeared probability distribution of events, whose shape depends on the quantum wavefunction of the gravitational state.

Finally,  we discuss how to approach the problem in terms of dressed, gauge invariant field operators~\cite{Donnelly:2015hta,Donnelly:2016rvo,Giddings:2018umg,Giddings:2019wmj}. Because of locality, our observers~\eqref{observers} are forced to interact with the field $\phi$ \emph{evaluated at their position}. 
So, instead of fixing the coordinates, one can build composite field operators containing a non local combination of the metric field. This is a gravitational Wilson line that has the moral purpose of prescribing at which position the field must be evaluated. In this context, different frames correspond to different dressings, though the transformation between frames at the operator level is not particularly enlightening. In Sec.~\ref{dressing} we show that under certain conditions it is possible to define dressed operators that are partially averaged over the gravitational degrees of freedom. Though these objects are of perhaps limited use, they have interesting transformation properties that are morally equivalent to the probabilistic coordinate transformation~\eqref{probability}. Despite being built out of standard scalar fields and being, naively, ``local" in any given frame, such operators transform with non-local convolutions under a change of frame.

In this paper \emph{frame} and \emph{localization} are used almost interchangeably. More precisely, a \emph{frame} is a set of coordinates that is completely gauge-fixed and we call \emph{localization} the procedure (i.e. identifying a set of observers) that allows one to define a \emph{frame}.

The paper is organized as follows. In Section \ref{sec:JTrev} we review a few aspects of JT gravity and its dual description. In Section \ref{JT} we discuss different frames in JT gravity, and the resulting quantum coordinate transformations and event smearing. In Section  \ref{prob} we give some ideas on how to define probabilistic coordinate transformations in quantum gravity, then apply these ideas to define probabilistic ``Lorentz boosts" in a model of linearized GR where the Hilbert space is restricted to a class of coherent states built from classical gravitational wave solutions. In Section \ref{dressing} we comment on the approach based on dressing local bulk operators. In Section \ref{sec:discussion} we give some concluding remarks. A few technical points are relegated to the Appendices.

{\bf Note added:} At around the same time of this pre-print's first appearance, a paper~\cite{Kabel:2024lzr} came out with similar conclusions about the localization of an event in the context of quantum reference frames (e.g.~\cite{Giacomini:2017zju}, see also~\cite{Torrieri:2022znj} for a related approach). Although the effect that we discuss here is entirely pinned on the fluctuations of the gravitational field---at least we work in this limit---the overlap between these approaches clearly deserves further study.

{\bf Notations:}
We use the \emph{mostly plus} convention for the metric.  Four dimensional coordinates have either Greek indices, $x^\mu$, or boldface character, $\bx$. Spatial coordinates have either Latin indices, $x^i$, or arrows, $\vec x$. With abuse of notation we sometimes  use $(t,x,y,z)$ instead of $(x^0,x^1,x^2,x^3)$.

\section{Jackiw-Teitelboim (JT) Gravity}  \label{sec:JTrev}
As we have discussed, the very definition of event is frame- (or observer-) dependent in quantum gravity. This remark seems fairly understood and uncontroversial, although only rarely has it been spelled out.

One  playground where this concept can be made particularly explicit is Jackiw-Teitelboim (JT) gravity, which is well known for its simplicity and connection to holography \cite{Almheiri2015-uf}.  In this context,  the existence of different  frames has been highlighted by Blommaert, Mertens and Verschelde (BMV)~\cite{Blommaert:2019hjr}, who calculated distances and correlators between bulk points in a specific frame that we are going to detail in the next section.

For later use we now give a brief review of the holographic Almheiri-Polchinski (AP) model \cite{Almheiri2015-uf} of Jackiw-Teitelboim (JT) gravity. The reader already familiar with the topic can skip directly to Sec.~\ref{JT}. One can also refer to the existing literature. The relationship between the particular Schwarzian action we will derive and the AP model of JT gravity was mentioned in \cite{Maldacena:2016hyu} and then realized independently in \cite{Engelsoy:2016xyb, Maldacena:2016upp, Jensen:2016pah}, though we will mostly follow \cite{Maldacena:2016upp} and the review article \cite{Mertens:2022irh}.

\subsection{A brief review of the model}

JT gravity is a two dimensional dilaton gravity theory, with action
\begin{align} \label{JT-action}
    S = \frac{1}{16 \pi G} \int d^2 x \sqrt{-g} (R+2/\ell^2) \phi + \frac{1}{8 \pi G} \int dx \sqrt{-\gamma} (K-1/\ell) \phi\, .
\end{align}
which can be shown to be a model of spherically symmetric perturbations to the near horizon geometry of a strongly charged black hole \cite{Nayak:2018qej} . The $G$ above is Newton's constant in two dimensions, and the parameter  $\ell$ sets  the radius of the (locally) AdS solutions.

In the AP model, we consider the gauge
\begin{align}
    ds^2 = e^{2\omega} (-dt^2 + dz^2)
\end{align}
where we set boundary conditions 
\begin{align} \label{JT-bc}
    g_{tt} = -\frac{\ell^2}{\delta^2} + \dots \;\;\;\;\;\;\;\; \phi = \frac{\ell \phi_b}{\delta} + \dots
\end{align}
with the boundary living at $z=\delta \rightarrow 0$. Variation of (\ref{JT-action}) with respect to the dilaton leads to $R=- 2/\ell^2$. In two-dimensions, this implies that our spacetime is locally $AdS_2$. In Poincaré coordinates, the metric is
\begin{align}
    ds^2 = \ell^2 \frac{-dT^2 + dZ^2}{Z^2}\, .
\end{align}
For later use, we also define a set of null coordinates $U = T+Z$, $V = T-Z$. 
Note, these are not necessarily the same coordinates used to set the boundary conditions, and in general $e^{2\omega} \not = \frac{\ell^2}{z^2}$. This will become important in a moment.

Normally, we would now vary the metric to get a (trivial) dynamics for the dilaton. However, since the dilaton appears linearly in the action, we are free to substitute the constraint $R=-2/\ell^2$ back into the action\footnote{At the quantum level, imposing the constraint manifests as evaluating the path integral $\int \mathcal{D} \phi e^{\int i \phi \times \text{constraint}}$ and producing a delta function $\delta(R+2/\ell^2)$ \cite{Maldacena:2016upp}. This delta function produces a Jacobian which changes the path integral measure to $\mathcal{D}f=\prod_t \frac{df(t)}{f'(t)}$}. This kills the bulk term, naively leaving no dynamics in the theory. There is, however, a subtlety, which is that we have not chosen which coordinate system the boundary conditions are set in. We are free to parameterize over this coordinate ambiguity, then substitute the parameterization into the remaining boundary action. The result will be a holographic theory of the ``parameterization degrees of freedom". Let us go into detail.

To parameterize over the coordinate ambiguity, we start by defining the surface of constant $z=\delta$ in the $(t,z)$ coordinates to live at $(T_b(t),Z_b(t))$ in Poincar\'e coordinates, and then take our boundary conditions (and therefore also the boundary action) to live on that surface. This appears to give us two functional degrees of freedom. However, the boundary conditions eliminate one of the degrees of freedom: in Poincar\'e coordinates, the metric boundary condition (\ref{JT-bc}) reads, to leading order:
\begin{align}
    g_{tt} = \ell^2 \frac{-\dot T_b(t)^2 + \dot Z_b(t)^2}{Z_b(t)^2} = -\frac{\ell^2}{\delta^2}
\end{align}
Noting that $Z=O(\delta)$, this equation fixes the $Z$ coordinate of the boundary to 
\be  \label{ZB}
Z_b(t) = \delta \, \dot T_b(t) + O(\delta^2).
\ee 
 The dilaton boundary condition does not fix any boundary degrees of freedom, so we are left with just $T_b(t)$ as the only boundary dynamical variable.

We are now ready to write our action in terms of these new degrees of freedom. The renormalized boundary action in the  $\delta \to 0$ limit becomes: 
\begin{align}\label{actionsch}
    S \ = \ \frac{1}{8\pi G} \int_{(T_b(t),\delta T'_b(t))} dx \sqrt{-\gamma} (K-1/\ell)\phi \ = \ - C \int dt \{T_b(t),t\} + O(\delta)
\end{align}
where $\{T_b(t),t\}$ is the Schwarzian derivative,
\begin{align}
    \{f(t),t\} = \frac{f'''(t)}{f'(t)} - \frac{3}{2} \left(\frac{f''(t)}{f'(t)} \right)^2
\end{align}
and $C=\frac{\phi_b}{32\pi G}$. At finite $\delta$, this can be thought of as a holographic theory of the embedding of the boundary into the bulk.

\subsection{Black holes, correlators and variances}
\begin{figure}[h]
\begin{center}
     \includegraphics[width=13cm]{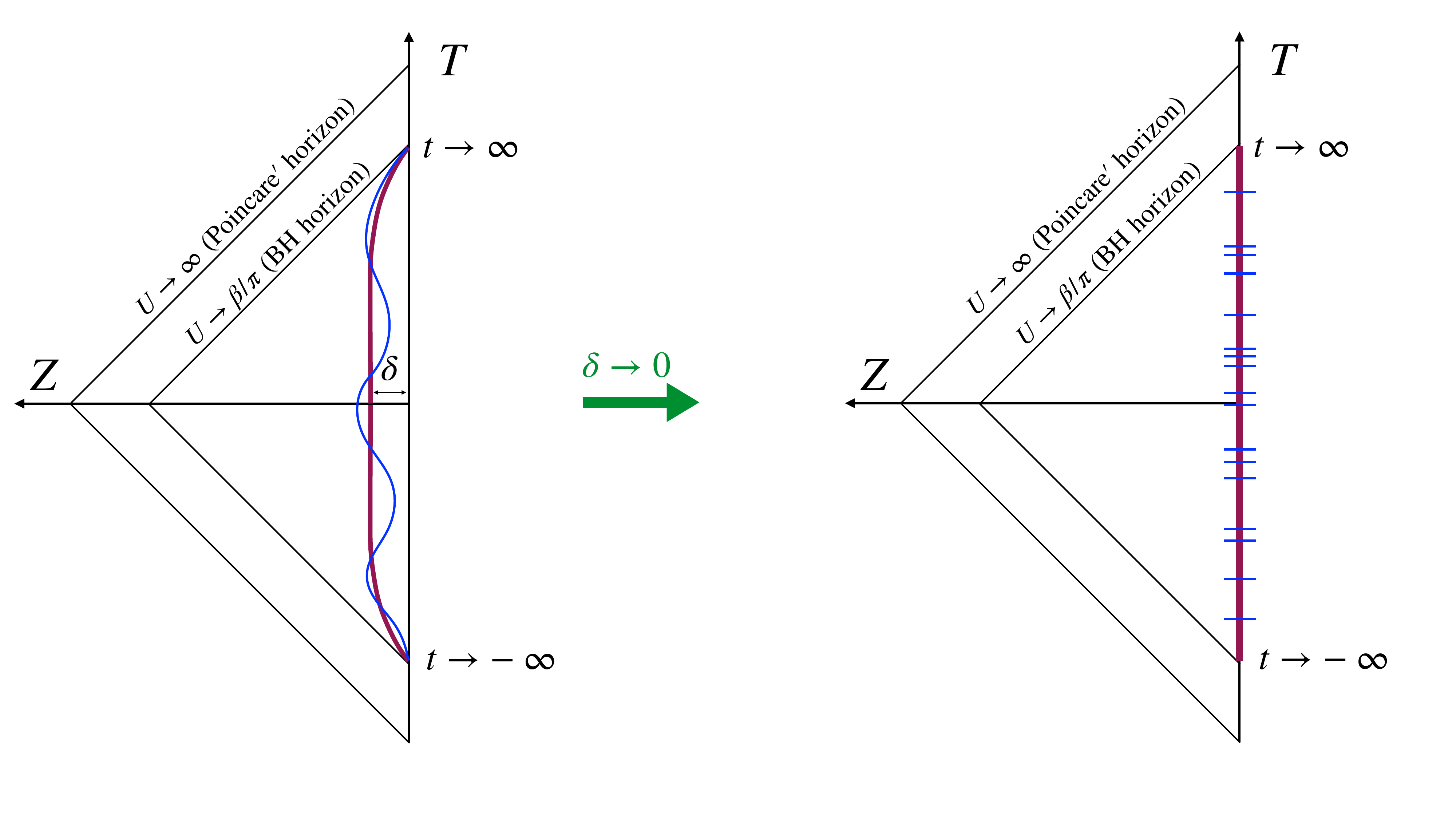}
  \end{center}
    \captionsetup{width=.95\linewidth}
\caption{\small A picture of the black hole solution in Poincar\'e coordinates. The full quantum state includes fluctuations  of the boundary  (in blue) around the saddle (in purple). We parameterize these fluctuations with the function $\epsilon(t)$. In the limit where the regulator $\delta$ goes to zero the boundary flattens into the $AdS$ boundary at infinity. The only surviving feature is that the time $t$ clicks at a different rate than the Poincar\'e time $T$.} \label{fig-2}
  \end{figure}

One can compute the response of the dilaton to the presence of a CFT in the bulk using the bulk equations of motion. Using the dilaton boundary conditions one can then relate the boundary particle configuration to the CFT stress tensor. For an ingoing impulse of energy $E=\frac{\pi^2}{\beta^2}$ one finds at $t>0$ the boundary configuration 
\cite{Mertens:2022irh}
\begin{align}
    T_{b,\text{BH}}(t) = \frac{\beta}{\pi} \tanh\left( \frac{\pi t}{\beta} \right) \, .
\end{align}
By extending the above to $t<0$ we get the ``eternal black hole" saddlepoint. While the boundary proper time $t$ goes to $\pm \infty$, the Poincar\'e time $T$ goes to $\pm \beta/\pi$. The resulting geometry in the bulk corresponds to a piece of $AdS_2$ bounded by $Z_{b,\text{BH}} (t) = \delta/ \cosh^2(\pi t/\beta)$ near the Poincar\'e boundary and by the two horizons $U = V = \beta/\pi$ (Fig.~\ref{fig-2}). 
We can perturb around this solution by defining $\epsilon(t)$ such that
\begin{align} \label{defeps}
    T_b(t) =   T_{b,\text{BH}}(t + \epsilon(t))\, ,
\end{align}
and expand the action~\eqref{actionsch} at quadratic order in $\epsilon$. This gives, up to total derivatives,
\begin{align} \label{e^2 action}
    S = \frac{C}{2} \int  dt \;\left[\ddot \epsilon^{\, 2} +\frac{4 \pi^2}{\beta^2} \dot \epsilon^{\, 2} \right]\, .
\end{align}
The presence of higher derivatives in the action is usually a worrisome signal of instabilities. Here, however, the unstable modes should be seen as pure gauge as explained in~\cite{Maldacena:2016upp} and should not be accounted for when calculating the propagator. By going to Fourier space and excluding the $n=0, \pm1$ gauge modes the Euclidean time propagator reads %$\tau$ as
\begin{equation}\label{series}
\bra \epsilon(u) \epsilon(0)\ket = \frac{2}{C}\sum_{n\not= 0,\pm1} \frac{e^{in u}}{n^2(n^2-1)} \, ,
\end{equation}  
where $u = - 2 \pi i t/\beta$. Different ways to calculate the above series are discussed in~\cite{Sarosi:2017ykf}. For $0<u<2\pi$ we get the known expression 
\begin{align}\label{feynmanprop}
    \braket{\epsilon(u)\epsilon(0)} =  \frac{1}{2 \pi C} \left(\frac{\beta}{2\pi}\right)^3  \Big[1 + \frac{\pi^2}{6} - \frac{1}{2}(u-\pi)^2  + \frac{5}{2} \cos u + (u-\pi) \sin u \Big] \, .
\end{align}
Mathematica also evaluates~\eqref{series} providing an analytic continuation of~\eqref{feynmanprop2} in terms of PolyLog functions, 
 \begin{equation}\label{feynmanprop2}
 \braket{\epsilon(u)\epsilon(0)} = 1 + \frac{5 \cos u}{2} -\text{Li}_2\left(e^{-i u}\right)-\text{Li}_2\left(e^{i u}\right)+i \left[\log \left(1-e^{-i u}\right)-\log \left(1-e^{i u}\right)\right] \sin u\, .
\end{equation}
As expected from thermal correlation functions, the above expression has branch cuts  along the Lorentzian time axis.

Any function $F$ of the Poincaré boundary coordinate  $T_b,Z_b$ becomes,  via equation (\ref{ZB}), a functional of $T_b(t)$ and thus,  via  (\ref{defeps}), a functional of the  fluctuating variable $\epsilon(t)$, whose quantum dynamics is governed  (at quadratic order) by the path integral with action (\ref{e^2 action}).  In App.~\ref{varianceAppendix} we find  the variance of such  a functional  to lowest order in $\epsilon$, 
\begin{align} %\label{simpto}
   \Delta F \equiv    \braket{F^2} - \braket{F}^2 = \left(\frac{d F}{dt}\right)^2 \braket{\epsilon^2} + \left( \frac{\partial F }{\partial \dot T_b } \, \dot T_{b,\text{BH}}\right)^2\braket{\dot \epsilon^2} \, ,
\end{align}
To obtain the above we have used $\braket{\epsilon}=\braket{\dot \epsilon}=0$, and the fact that in this theory $\braket{\epsilon(t)\dot \epsilon(t)}=0$. At quadratic order we are left with $\bra \epsilon^2\ket$ and  $\bra \dot \epsilon^2\ket$. The first can be obtained by simply evaluating one of the two correlators above in the coincident limit.  
\begin{equation} \label{varianceeps}
\bra \epsilon^2\ket\ =\ \left(\frac72 - \frac{\pi^2}{3}\right) \frac{\beta^3}{(2 \pi)^4 C} %\equiv \alpha \, \frac{\beta^3}{C}
\end{equation}
As for the second quantity of interest, 
\begin{align} 
\bra \dot \epsilon^2\ket \ = \ \lim_{t_1\rightarrow t_2} \frac{d}{dt_1} \frac{d}{dt_2} \bra \epsilon(t_1)  \epsilon(t_2)\ket =  - \lim_{t \rightarrow 0} \ddot G(t), 
\end{align}
with $G(t_1 - t_2) = \bra \epsilon(t_1)  \epsilon(t_2)\ket$ the Wightman function. By Wick rotating~\eqref{feynmanprop2} to Lorentzian time we obtain instead the time-ordered propagator, $G_F(i \beta u/2 \pi) = \bra \dot \epsilon^2\ket$. From the relation $G_F(t_1 - t_2) = \theta(t_1 - t_2) G(t_1 - t_2) + \theta(t_2 - t_1) G(t_2 - t_1)$ we also obtain $\ddot G(t) = - \ddot G_F(t)$.  So by evaluating the second derivative of~\eqref{feynmanprop2} in the coincidence limit we finally get
\begin{equation} 
\braket{\dot \epsilon^2}  = \frac{3 \beta}{8 \pi^2 C}\, .
\end{equation}
The variance of a generic functional $F$ then reads:
\begin{align}\label{varianceEquation}
      \Delta F = \frac{\beta}{4 \pi^2 C} \left[\frac32  \left( \frac{\partial F }{\partial \dot T_b } \dot T_{b,\text{BH}} \right)^2 + \, \frac{21 - 2 \pi^2}{24 \pi^2} \left(\frac{d F}{dt}\right)^2 \beta^2\right]\, .
\end{align}
The variance of $F$ in the full non-linear theory will receive corrections to this expression in higher orders of $\beta/C$, which is the loop expansion parameter of JT gravity when expanding around the black hole saddlepoint. The appearance of the loop parameter in front of~\eqref{varianceEquation} highlights the quantum origin of the effect we are discussing.

In the next  section we will be interested in evaluating the variance of various bulk ``events'' which are sharp in Poincar\'e coordinates, but become smeared when expressed in a coordinate system  anchored  to the bulk\footnote{We will give several examples of such coordinate systems.}, i.e. expressed in terms of $\epsilon$: the Poincar\'e coordinates $T,Z$  of the event will become functionals of  $\epsilon$, and therefore subject to an uncertainty computed by the general formula (\ref{varianceEquation}). As an example, we will be  especially interested in evaluating the variance of the location of the future horizon, which normally lives at $U=V= \beta/\pi$ in Poincar\'e coordinates, but can be more geometrically identified as the past lightone of the event where the boundary intersects the singularity (as can be easily seen in \cite{Almheiri2015-uf}), or where the boundary location $Z_b(t) \rightarrow 0$ .

\section{Frames and horizon smearing in JT gravity}\label{JT}

In this section we discuss the construction of different {\em localizations} in JT gravty. By localization, we mean a way of constructing a local coordinate system (bulk frame) starting from the boundary. This leads to  an explicit realisation of  the ideas outlined in Sec.~\ref{sec:intro1} .

As in other holographic models, a bulk point in JT gravity can be labeled in a gauge invariant way by anchoring/\emph{localizing} it to the boundary. The localization can be done in different ways, which leads to inequivalent labelings of bulk events. Two different  localizations will generally be related by a functional of $T_b(t)$---the fluctuating degree of freedom living on the boundary, as reviewed in the previous section.  Since $T_b(t)$ is fluctuating, a point with specific spacetime coordinates in one localization will have a superposition of coordinates in a different localization.
It is in principle possible to find this superposition of labels explicitly, but this  is cumbersome when considering the full boundary theory.  However calculations become relatively simple if we restrict ourselves to  the linearized boundary theory,  i.e. the Gaussian path integral over the fluctuation $\epsilon(t)$ defined in (\ref{defeps}): the calculation  simply amounts to finding the average and variance of the appropriate label distribution.

In this section we carry out this computation for the location of the  horizon of an eternal black hole (as defined in Poincar\'e coordinates) as it appears in various localizations, which  are schematically   summarized in Table~\ref{tab}, and will be described in more detail in the following subsections.  
 As we will see,  in certain localizations the horizon is a sharp surface (as in the  particular case of the localization defined in \cite{Blommaert:2019hjr}) but in others  it appears \emph{smeared}. 
\begin{table}[h]%\vspace{-1cm}
\begin{tabular}{|c|c|}
\hline
 \begin{tabular}{c}
  {\bf Poincar\'e coordinates} $(U,V)$ or $(T,Z)$ \\[2mm]
  Locally, the geometry in JT gravity is that of $AdS_2$. \\ 
  As such, standard  Poincar\`e coordinates constitute a well \\
  defined frame, which however is not defined operationally, \\
  \emph{i.e.} in terms of  its relation with the boundary. 
  \end{tabular} &
    \begin{tabular}{c}
   \includegraphics[width=1.4in]{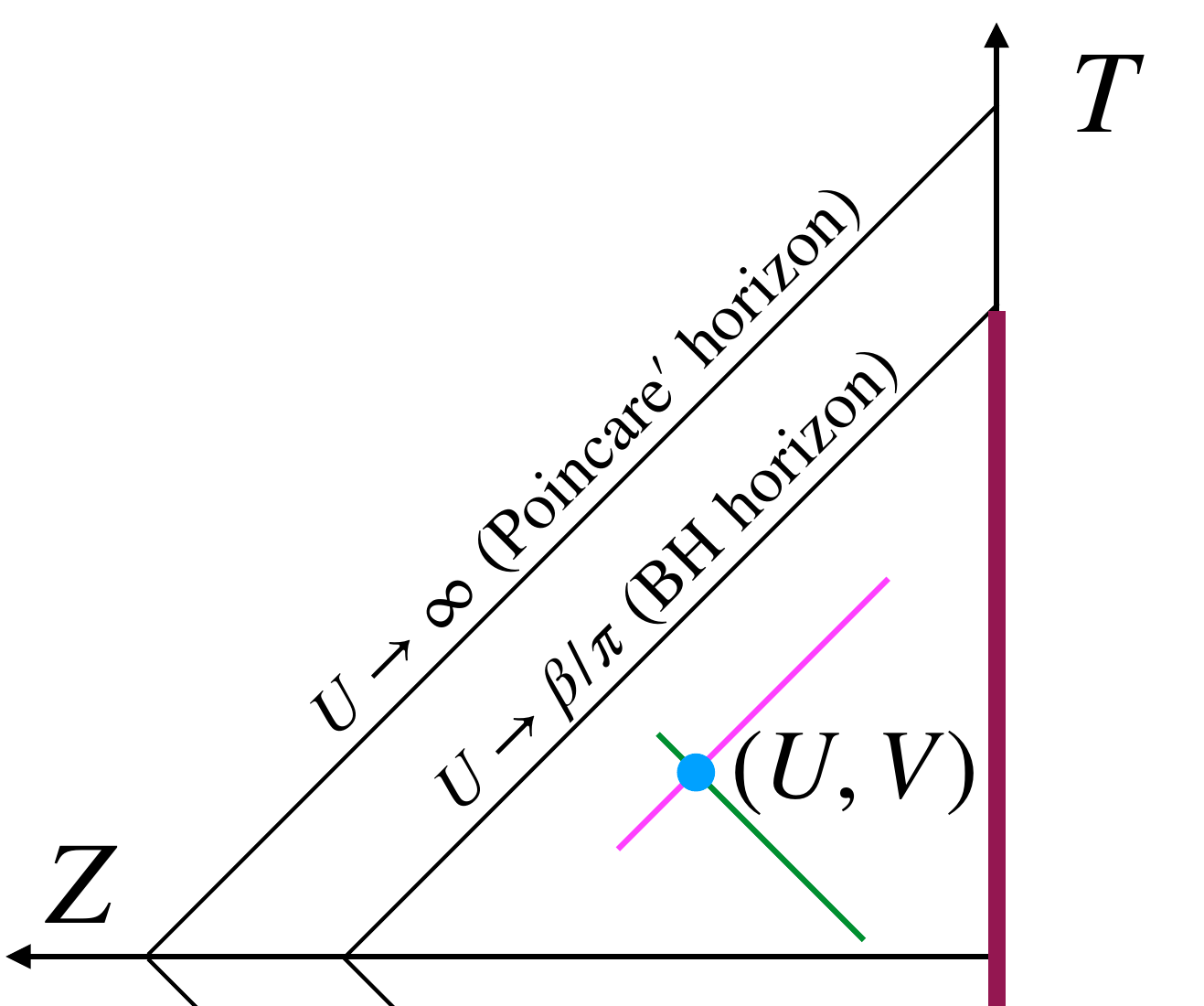} \end{tabular}
  \\
  \hline
    \begin{tabular}{c}
 {\bf BMV localization} $(t_1,t_2)$ \\[2mm]
  Blommaert, Mertens and Verschelde~\cite{Blommaert:2019hjr} have proposed \\
  to label a point in the bulk by sending in a light ray \\ 
  from boundary time $t_1$ and another light ray \\ 
  backward in time from time $t_2$. 
    \end{tabular} &
    \begin{tabular}{c}
   \includegraphics[width=1.4in]{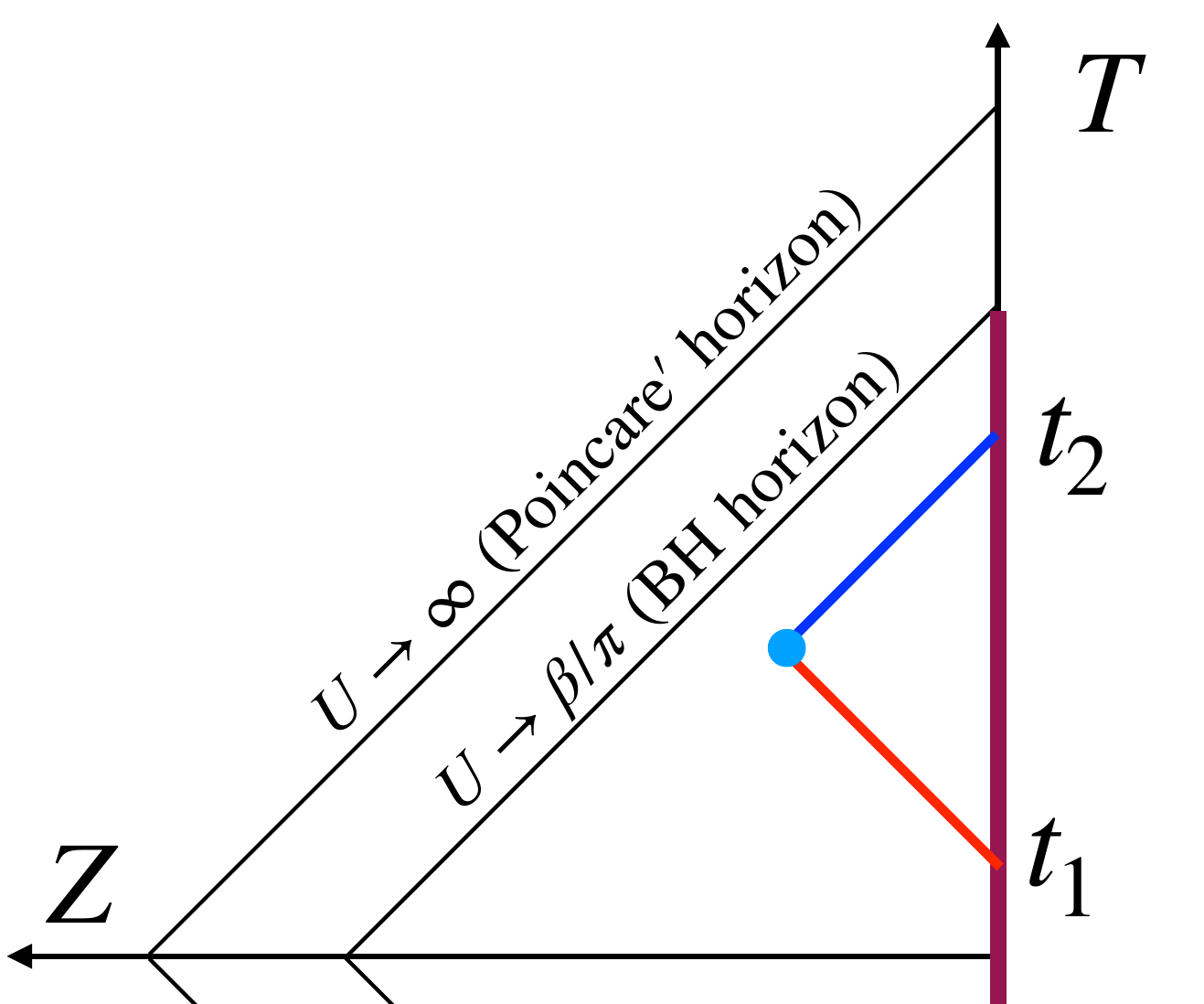} \end{tabular}
  \\
  \hline
    \begin{tabular}{c}
   {\bf Spacelike localization} $(t ,\bar L)$ \\[2mm]
  Standard procedure in holography. We send spacelike \\ geodesics  
  orthogonally to the boundary for a (renormalized) \\
  proper length $\bar L$.  In the $\delta \rightarrow 0$ limit this is equivalent \\ 
  to sending geodesics at constant Poincar\'e time $T$ 
  \end{tabular} &
    \begin{tabular}{c}
   \includegraphics[width=1.4in]{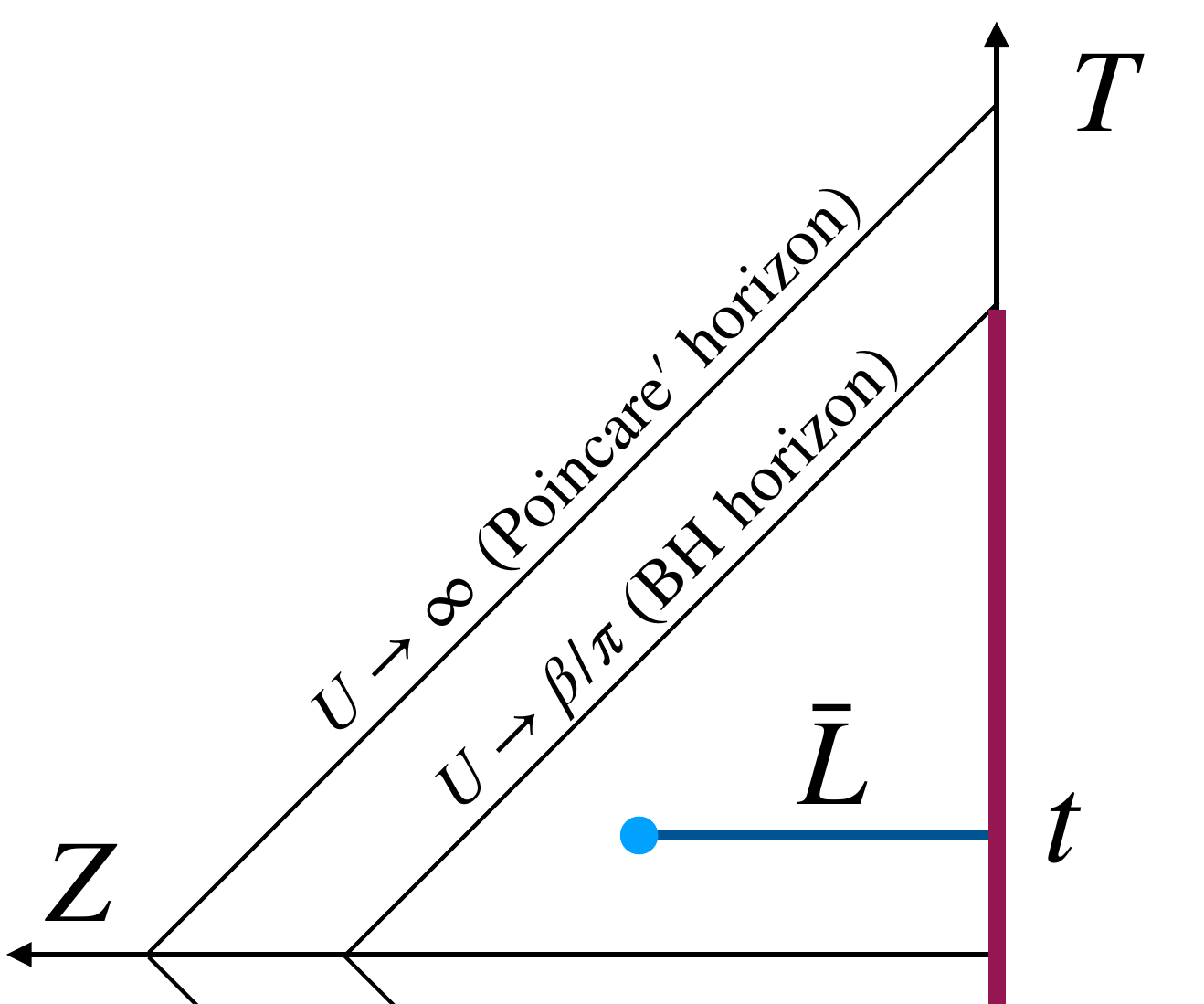} \end{tabular}
  \\
  \hline
    \begin{tabular}{c}
 {\bf Timelike localization} $(t ,\tau)$ \\[2mm]
  We lean inside the bulk orthogonally to the boundary\\
  for a fixed renormalized length and then send a timelike\\
  geodesic in the orthogonal direction for a proper time $\tau$.
   \end{tabular} &
    \begin{tabular}{c}
   \includegraphics[width=1.4in]{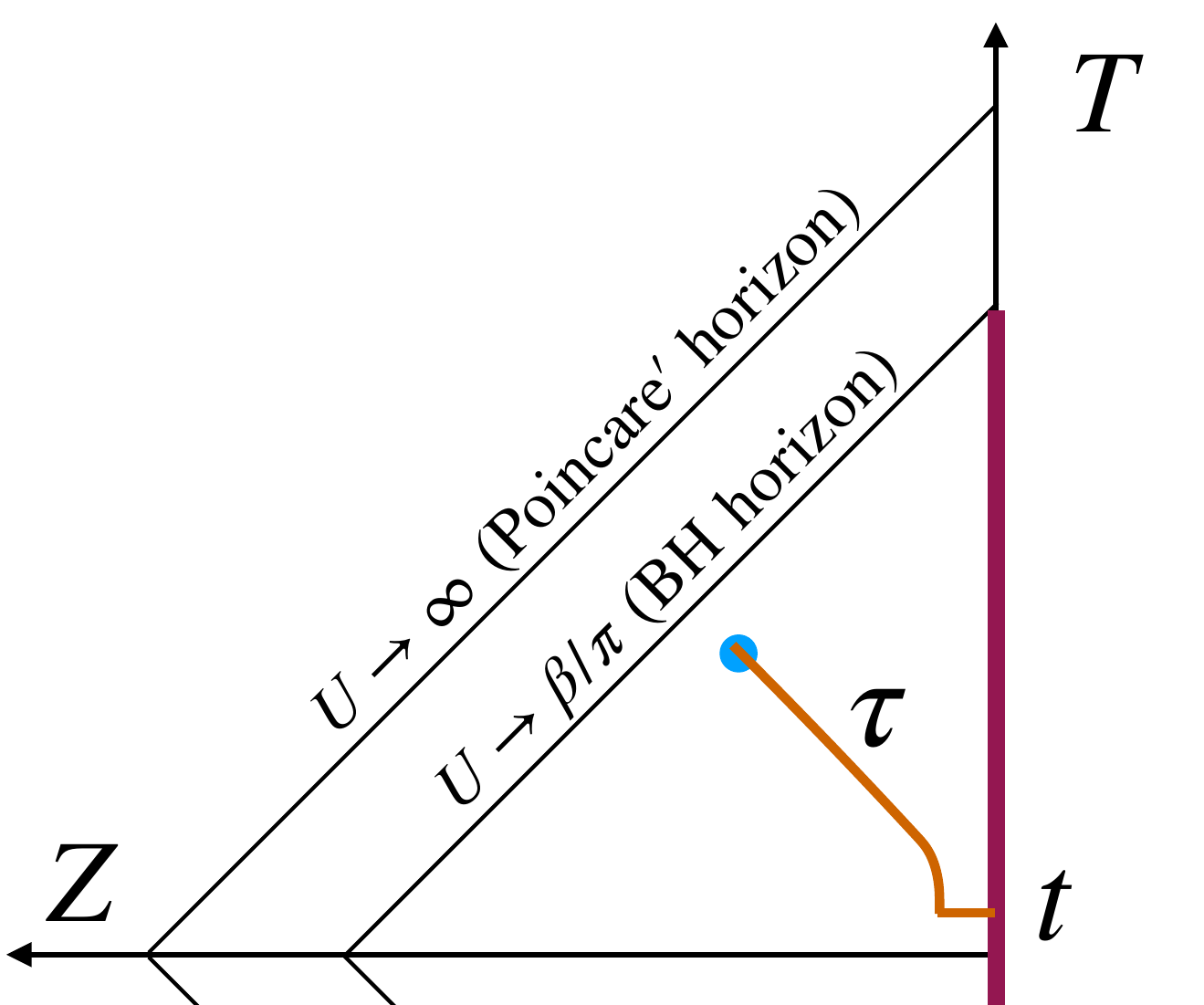} \end{tabular}
  \\
  \hline
\end{tabular}  \captionsetup{width=.95\linewidth}
\caption{\small The localizations considered in this paper.} \label{tab}
\end{table}

\subsection{BMV Localization}\label{BMV_localization_section}

The first localization we will discuss was already explored in~\cite{Blommaert:2019hjr}:  we send a light ray out to the event of interest starting from boundary time $t_1$ and another light ray backward in time from a later boundary time $t_2$.  The intersection of these light rays defines an event labeled $(t_1,t_2)$.   Some simple geometric reasoning reveals that the Poincar\'e coordinates of the event of interest are given by
\begin{align}
    U = T_b(t_2)\label{U}\, ,
    \\
    V = T_b(t_1)\, .
\end{align}
The bulk AdS$_2$ metric in the coordinates $t_1,t_2$  is\footnote{We can take the  boundary regulator $\delta\to 0$ without need of a renormalization procedure. This will not be the case for the other localizations which we will discuss later, which require a subtraction scheme.}
\begin{align} \label{metricBMV}
    ds^2 = \frac{-4 \ell^2 \dot T_b(t_1)\dot T_b(t_2)}{(T_b(t_1) - T_b(t_2))^2}dt_1 dt_2\, .
\end{align}

We see that $t_1$ and $t_2$ are null coordinates in this localization. The fact that the  conformal factor contains the fluctuating variable $T_b$ indicates that the distance between two events will also fluctuate in general. However, two BMV events that share the same value of $t_1$, or $t_2$, are always, by construction, at exact null separation. Accordingly, the BMV localization preserves a sharp lightcone structure, outside of which the commutator of a test field strictly vanishes~\cite{Blommaert:2019hjr}. For all other frames discussed below the statement of two events being at null separation is instead probabilistic.

\begin{figure}[h]
%\vspace{-1cm}
\begin{center}
     \includegraphics[width=9cm]{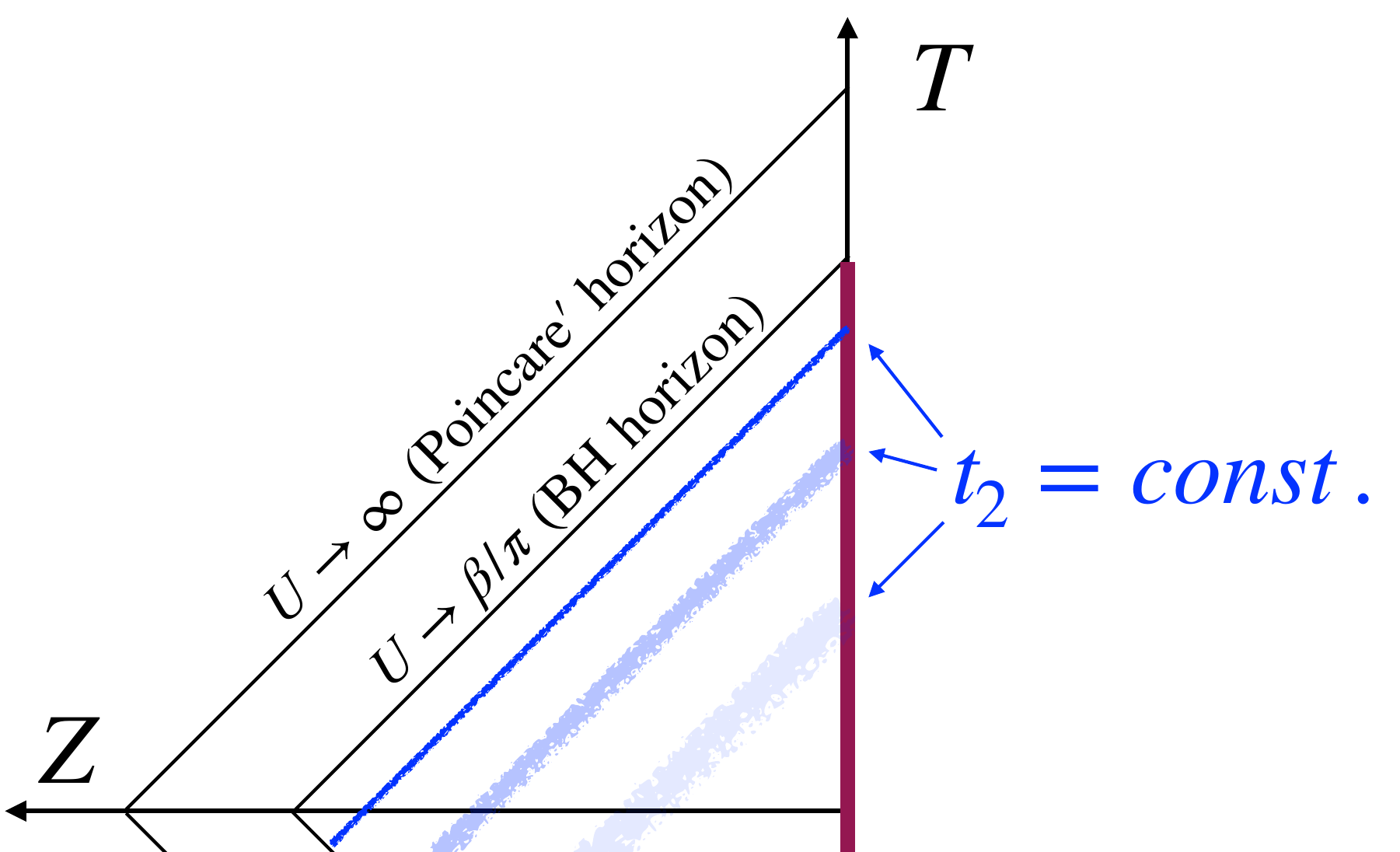}
  \end{center}
    \captionsetup{width=.95\linewidth}
\caption{\small In the Poincar\'e frame we plot events that are defined as $t_2 = const.$ in the BMV localization. The smearing $\Delta U$ of the corresponding curves shrinks as we approach the BH horizon. } \label{fig-3}
  \end{figure}

\paragraph{Smearing of Poincar\'e coordinates}
Despite respecting the classical lightcone structure, the BMV frame clearly differs from the $AdS_2$ Poincar\'e coordinates, in that an event specified by  $t_1,t_2$ in the BMV frame has some \emph{probability amplitude} of being described by  some $U,V$ in Poincar\'e coordinates.   
The relative smearing
is quantified by the variance $\Delta U$  of the coordinate $U$ at given $t_2$, which we can obtain from ~\eqref{U} and~\eqref{varianceEquation}:
\begin{align} \label{corollaryto}
\Delta U \ = \ \dot T^2_\text{b,BH}(t_2) \braket{\epsilon^2}\  = \ \frac{21 - 2 \pi^2}{6 (2 \pi)^4}\frac{\beta^3}{ C \cosh^4\frac{\pi t_2}{\beta}}\, .
\end{align}
and similarly for $\Delta V$ and $t_1$. The smearing of the $t_2 = constant$ surfaces as perceived in the Poincar\'e frame is schematically depicted in Fig.~\ref{fig-3}. 
Since connected correlation functions are negligible at orders where interactions can be ignored, $\epsilon(t)$ can be accurately treated as a Gaussian-distributed variable, and we can interpret the above as the fact that the coordinate transformation from the BMV frame to the Poincaré frame has a Gaussian probability distribution whose width is given by equation (\ref{corollaryto}).

\paragraph{Horizon Smearing}
We can also consider what happens to the horizon of the bulk black hole as seen in the BMV frame. The horizon can be defined as the null ray that intersects the boundary at $Z_b=0$. In the black hole configuration, this intersection point lives at $t\rightarrow \infty$. However, the fluctuations of $T_b(t)$ go to zero in this limit. This can be seen as follows. The variance of $\epsilon$~\eqref{varianceeps} is time independent. From the definition~\eqref{defeps}, the fluctuations in $T_b(t)$ and $\epsilon$ are related by derivatives of $T_\text{b,BH}$, which rapidly flattens out in the $t\rightarrow \infty$ limit. 
The fact that the black hole horizon is sharp in the BMV frame can also be seen as a corollary to~\eqref{corollaryto}: the variance of $U$ vanishes as $t\rightarrow \infty$ and the black hole horizon is located at a precise value $U = \beta/\pi$ and  $t_2\rightarrow \infty$. In summary, \emph{the BMV  localization does not result in  any smearing of the horizon. }

\subsection{Spacelike Localization}\label{sectionSpacelikeLoc}

If we want to convey the point of view of the observers falling into the black hole we should use their proper time. This requires some form of renormalization as the proper length to the boundary becomes infinite in the $\delta\rightarrow 0$ limit.  Note that the BMV localization did not require this since it only refers to the boundary time, and is therefore insensitive to the $\delta\rightarrow 0$ limit.

As a warm up, we first address the \emph{spacelike localization}: at some given boundary time $t$ we send a spacelike geodesic orthogonal to the boundary and measure its proper length $L$, labeling the endpoint of the geodesic $(t,L)$. This is a rather standard procedure in holography (e.g.~\cite{Papadodimas:2015jra,Donnelly:2015hta,Almheiri:2017fbd}), although, to the best of our knowledge, the simple fact that the position of the horizon can become indefinite has never been highlighted.

Points labeled by $(t,L)$ in this localization correspond to\footnote{This looks like a geodesic at constant Poincar\'e time $T$, which would be operationally ill defined. However, what we are actually doing is sending geodesics orthogonally to the boundary, with no reference to Poincar\'e coordinates.  In App.~\ref{appendixPerpLoc} we show that this prescription leads to~\eqref{spacelikeGeo} and~\eqref{spacelikeGeo2}  at leading order in $\delta$.}  
\begin{align}\label{spacelikeGeo}
    T &= T_b(t)\, ,
    \\
    Z &= Z_b(t) e^{L/\ell}\, \label{spacelikeGeo2}.
\end{align}
in Poincar\'e coordinates, where $L$ is the geodesic length from the boundary. This quantity diverges in the $\delta \rightarrow 0$ limit, although \emph{differences} in $L$ are finite.  We can eliminate  the overall divergence by defining a renormalized distance $ \bar L$:
\begin{align}
    \bar L = L + \ell \ \log  \frac{\delta}{\zeta}\, , 
\end{align}
where, a little pedantically, we have introduced the renormalization scale  $\zeta$ to maintain dimensional consistency. $\zeta$ obviously makes $\bar L$ scheme dependent. We can think of the family of schemes labeled by $\zeta$ as a family of localizations with the property that all members of the family agree on differences in $\bar L$, but not on the value of $\bar L$ itself. Finite,  scheme-independent quantities are those which depend on $\bar{L}$ and $\zeta$ through the combination $\zeta e^{\bar{L}/\ell}$.

\paragraph{Horizon Smearing}
We now compute the variance in the geodesic distance which a spacelike observer leaving at time $t$ must traverse before they reach the future horizon\footnote{Note that the observer won't be able to see when they cross the horizon, but they will be able to use lightrays to continuously communicate their current clocktime/propertime to another observer at infinity. The last clocktime report the observer at infinity sees (which they will see when $t \rightarrow \infty$, i.e. when $Z_b \rightarrow 0$)  will contain the proper time at which the infalling observer reached the horizon. As discussed at the end of the previous subsection, since we are implicitly only considering a space of states perturbatively close to a black hole of temperature $\beta$, and because the bulk geometry is fixed to AdS$_2$, this will always occur at the same outgoing Poincar\'e time $U_\text{hor}=T+Z=\beta/\pi$}. Using $U_\text{hor}=T+Z=\beta/\pi$ and inverting \eqref{spacelikeGeo} we obtain: 
\begin{align} \label{renormL}
     \bar L_\text{hor} =  \ell \log \left(\frac{\frac{\beta}{ \pi} - T_b}{\zeta \, \dot T_b}\right)  
\end{align}

The fact that the renormalization in (\ref{renormL}) is by an additive constant has the important consequence that variances in $ \bar L$ are independent of $\zeta$. 
Using \eqref{varianceEquation} we obtain
\begin{align}
\Delta \bar L_\text{hor} = \frac{\beta}{4 \pi^2 C} \left[\frac32  + \frac{21 - 2 \pi^2}{6\left( 1 + e^{2 \pi t/\beta}\right)^2}\right]\, .
\end{align}
This is independent of $\zeta$, and converges to a non-vanishing constant for $t \rightarrow \infty$.

\paragraph{Metric}
The metric in this localization is
\begin{align} \label{spacelikemetric}
    ds^2 = d \bar L^2 +  \frac{2 \ell \ddot T_b}{\dot T_b} dt d\bar L + \left(\ell^2 \frac{\ddot T_b^2}{\dot T_b^2}  - e^{-2 \bar L/\ell} \right) dt^2 
\end{align}
where  $\zeta$ has been set to $\ell$.

\subsection{Timelike Localization}
As discussed earlier, it would be desirable to have a localization in terms of families of physical observers, but neither the BMV nor the spacelike localization are of this type. In this subsection  we give a proposal for constructing such a physical frame.

A concrete way of defining a timelike frame in JT gravity proceeds as follows:  we send observers out from the  (regulated)  boundary at  each boundary time $t$,  with an initial velocity parallel to the boundary at that time. % (i.e. with its same velocity).
When the observer reaches an event in the bulk, they read the proper time on their clock $\tau$, and label the event $(\tau,t)$. 
This localization is more ``physical'' in that the ftimelike geodesics could be the trajectories of real observers.

As we show in App.~\ref{tangentialLoc}, in the limit where the regulator $\delta \to 0$, the prescription described above gives the same family of geodesics (up to $O(\delta)$) as the simplified prescription in which the observers start from the boundary at rest with respect to Poincar\'e coordinates, i.e. with the initial condition ${dZ\over d\tau}(\tau=0) =0$. Describing the geodesic paths of these observers gives a simple  relation between the $(\tau,t)$ labels and the Poincar\'e coordinates $T,Z$ in terms of the  boundary degree of freedom $T_b(t)$:
\begin{align}\label{timelikeGeo1}
    T(\tau,t) &= T_b(t) + Z_b(t) \tan (\tau/\ell)\, ,
    \\\label{timelikeGeo2}
    Z(\tau,t) &= \frac{Z_b(t)}{\cos (\tau/\ell)}\, .
\end{align}
where $Z_b(t)=\delta \dot T_b(t)$ as in equation (\ref{ZB}).

 One drawback of a ``realistic" localization such as this one is that now $\delta$ appears multiplicatively in $Z_b$ and, in the limit $\delta \to 0$, the proper time always approaches the same constant $\tau \to \pi \ell/2$ for any finite $T,Z$. In order to obtain a faithful labeling of spacetime points from these timelike geodesics one would need to extract the subleading term in the $\delta$ expansion\footnote{For the curious reader, the procedure is as follows. Inverting \eqref{timelikeGeo1} and \eqref{timelikeGeo2} gives
\begin{align} \label{deltalinear}
    \tau  = \frac{\pi \ell}{2} -  \frac{2 \ell  \dot T_b}{U - T_b} \, \delta + O(\delta^2)
\end{align}
to leading order in the boundary regulator. This has a finite universal part, and a vanishing piece that depends on the $U$ coordinate. The most natural renormalization procedure with a non-trivial result thus consists of affinely reparameterizing $\tau \rightarrow (\ell/\delta)(\pi \ell/2 - \tau)$.  This picks up the linear term in $\delta$ in~\eqref{deltalinear}. }.

As an alternative, we can get around this issue with a modification to the localization: instead of launching the observers directly from the boundary, we instead launch them from the end of a spacelike geodesic and orthogonally to it. The spacelike geodesic is of the same type as those discussed in the previous subsection. It starts at $Z_b$ and sticks into the bulk a renormalized  distance $\bar L$. This is represented in the bottom panel of Table \ref{tab}.  The path described by the  observers is  now:
\begin{align}\label{timelikeGeo1.2}
    T(\tau,t) &= T_b(t) + \zeta \dot T_b(t)e^{\bar L/\ell}   \tan (\tau/\ell)\, ,
    \\\label{timelikeGeo2.2}
    Z(\tau,t) &=  \frac{\zeta \dot T_b(t)e^{\bar L/\ell}  }{\cos (\tau/\ell)}\, .
\end{align}
and we are free to make the (convenient) choice $\bar L = 0$.

Morally, this renormalization procedure is the same as for spacelike geodesics, and $\zeta$ can again be thought of as parameterizing a family of localizations. A major difference from the spacelike case, however, is that differences in proper time are no longer $\zeta$-independent.

With regards to its operational interpretation, this localization is again only well defined insofar as it is the limit of a localization where observers leave the boundary without reference to the embedding of the boundary in Poincar\'e coordinates. Since this localization starts off with observers moving a spacelike distance into the bulk, our discussion at the bottom of section \ref{sectionSpacelikeLoc} guarantees that it is well defined.

\paragraph{Horizon Smearing} We will now compute the variance in proper time it takes for an observer leaving the boundary at time $t$ to reach the future horizon, which once again lives at $U=\beta/\pi$. Inverting \eqref{timelikeGeo1.2} and \eqref{timelikeGeo2.2} gives
\begin{align}
    \tau_\text{hor} = \ell \arccos{\frac{2\gamma/\zeta}{1 + \gamma^2/\zeta^2}}\, , 
\end{align}
where
\begin{align}
    \gamma \equiv \frac{\beta / \pi - T_b}{\dot T_b} \, ,
\end{align}
which, at the black hole saddlepoint
\begin{align}
    \gamma_\text{BH} = \frac{\beta}{2 \pi} \left( 1 + e^{-\frac{2\pi t}{\beta}} \right)\, .
\end{align}
Using \eqref{varianceEquation} we obtain 
\begin{align}
 \Delta \tau_{\rm hor} &= \left( \frac{2 \ell \zeta}{\gamma_\text{BH}^2 + \zeta^2} \right)^2 \left[  \left(\frac{\gamma_\text{BH} \ddot T_b}{\dot T_b} + 1 \right)^2 \braket{\epsilon^2} + \gamma_\text{BH}^2 \braket{\dot \epsilon^2} \right]
 \\
 &= \frac{\beta}{8 \pi^2 C} \left( \frac{2 \ell \zeta}{\gamma_\text{BH}^2 + \zeta^2} \right)^2 \left[ \frac{\beta^2}{\pi^2} \frac{21 - 2\pi^2}{12} e^{-\frac{4 \pi t}{\beta}}  + 3 \gamma_\text{BH}^2  \right]\, .
\end{align}

This converges very quickly to a constant as $t \rightarrow \infty$, converging quicker for hotter black holes, and to a lower value.

The parameter $\zeta$  encodes the renormalization-scheme dependence, which physically indicates how far from the true AdS boundary we release the free-falling observers.

\paragraph{Metric} The metric in this localization is (with $\zeta = \ell$)
\begin{align}
    ds^2 = -d\tau^2 - 2 \, d\tau dt   + \left[\left(\frac{\ell \, \ddot T_b}{\dot T_b} \right)^2 -\left(\cos (\tau/\ell) + \frac{\ell\, \ddot T_b \sin(\tau/\ell)}{\dot T_b}\right)^2\right] dt^2\, .
\end{align}
We can also synchronize the observers' clocks differently by defining the retarded time $\tilde \tau = \tau + t$. The metric then appears in the standard ``synchronous gauge" form 
\begin{align}
    ds^2 = -d\tilde \tau^2   + \left[1+ \left(\frac{\ell \, \ddot T_b}{\dot T_b} \right)^2 -\left(\cos (\tau/\ell) + \frac{\ell\, \ddot T_b \sin(\tau/\ell)}{\dot T_b}\right)^2\right] dt^2 \, .
\end{align}

\subsection{Localizations, regulators and the second boundary of $AdS_2$}

Notice that some of the localizations required some kind of renormalization procedure, due to the fact that the affine parameter along the geodesics degenerates as we approach the boundary. 
This generally came at the cost of introducing a scheme dependence through a parameter $\zeta$. We chose to interpret the family of schemes parameterized by $\zeta$ as a family of localizations.

Interestingly, scheme dependence was not an issue in the BMV localization. This is because the BMV localization only references boundary times, which survive trivially in the $\delta\rightarrow 0$ limit. Indeed, insofar as the regulator is an unphysical mathematical convenience, one might argue that the only well defined localizations are those that are able to make it back to the boundary, since every other localization would require some reference to the affine parameter of the observers. This implies that referencing the black hole interior in a scheme independent way requires the second boundary. These geometric difficulties with one sided interior reconstruction are not unfamiliar in AdS/CFT.

\section{Probabilistic coordinate transformations} \label{prob}
  
One of  the most striking   features  which emerges from the previous section is that, in JT quantum gravity, the  bulk location of an event becomes  probabilistic under a coordinate transformation. Consider for definiteness an event labeled by $(t_1,t_2)$ in the BMV frame of section 3.1.  For a bulk observer using Poincar\'e coordinates $(U,V)$, the values of the coordinates of the event will be given  by a probability distribution which (in the quadratic approximation we used) is essentially a Gaussian centered around the classical coordinates $U(t_2), V(t_1) $ with a variance given by equation (\ref{corollaryto}). In other words, changing from BMV to Poincar\'e coordinates has introduced a probabilistic component in the labeling of the event.

We can go one step further, and  repeat the same argument for the coordinate transformation connecting  a pair of  bulk-anchored frames (e.g. the BMV frame  and one of the time-like frames, or two different time-like frames constructed from two different families of infalling observers): an event with coordinates $(t_1,t_2)$ in the BMV frame will now have a  probability distribution of coordinates $(t,\tau)$ as seen from, say, the time-like frame (\ref{timelikeGeo1.2}),(\ref{timelikeGeo2.2}), with a variance which can in principle be calculated by first computing the explicit coordinate transformation
\be \label{ct1}
t_1,t_2 \ \longrightarrow \ t(t_1,t_2), \ \tau(t_1,t_2)\ ,
\ee
expanding it in $\epsilon$ up to quadratic order, and computing the variances $(\Delta t)^2, (\Delta \tau)^2, \Delta t \Delta \tau$  using equation. (\ref{feynmanprop}).  Notice that in general,  (unlike in the calculation of (\ref{corollaryto})) the unequal-time correlation function will enter the result.

This is  an explicit example of what we called in the introduction a {\em probabilistic coordinate transformation:} given two frames defined by two sets of  observers A(lice) and B(ob), and a quantum state of the gravitational system, asking ``what is  the probabilistic coordinate transformation relating the two frames $\bx_\A$ and $\bx_\B$" means asking ``what is the probability distribution for an event, labeled  $\bx_\B$ in Bob's frame, to be labeled $\bx_\A$ in Alice's frame". We denote this probability distribution by
\be \label{ct2}
dP_{\bx_\B \to \bx_\A} = p(\bx_\A | \bx_\B) d^d \bx_\A, 
\ee
In $d=2$, JT gravity provided  us with a way of computing this probability distribution explicitly (at least order by order in  perturbation theory around a classical spacetime, e.g. the BH) in terms of the Schwarzian path integral.

More generally,  this probability   can in principle be  expressed in terms of the quantum gravitational state density matrix $\rho_g$ by
\be \label{ct3}
p(\bx_\A | \bx_\B) = Tr \left(\rho_g \,  Q_{\bx_\B \to \bx_\A}\right)\, ,
\ee
where $Q_{\bx_\B \to \bx_\A}$ is an appropriate projection operator. Although the latter is very hard to even define in general, and we will not attempt to do it here. One can nevertheless gather some intuition on how to construct it at least  in a semiclassical context, as  we will speculate in the next subsection. We will then carry out an explicit calculation  in a  linearized gravity example in subsection \ref{boost},  where the state is a superposition of gravitational coherent  states.

We stress that the probabilistic nature of the coordinate transformation arises because the gravitational field is in superposition, {\em not} the observers, whose coordinates  are considered  classical  variables.

\subsection{The use of coherent/semiclassical states}

We will now try to make sense of probabilistic coordinate transformations in a more realistic context. To be definite, we consider Einstein gravity in $d$ spacetime dimensions, with frames (classically) defined by families of freefalling observers.

In order to make the calculation of the probability distribution (\ref{ct3}) more approachable in the quantum version of the theory, we make the following assumption:

\begin{quote}
\emph{there exists a set of semiclassical  states, labeled by  $|g\ket$ on which the coordinate transformation between Alice and Bob is a standard (deterministic) one}. \\
\begin{equation}
\bx_\B \ \longrightarrow \ \bx (\bx_\B;g)\, .
\end{equation}
\end{quote}
In the above, $g$ refers to the metric of the semiclassical  state $|g\ket$, and  $\bx (\bx_\B;g)$ are the coordinate of Bob-like observers as measured in the Alice frame. They depend on the classical metric $g$, as the latter affects the trajectories of different families of geodesic observers.  Effectively, on each semiclassical state $|g\ket$, the coordinate transformation-probability is delta-like in the classical limit
\begin{equation}\label{definiteFrameChange}
p(\bx_\A|\bx_\B) \sim \delta^4(\bx_\A - \bx(\bx_\B;g))\, .
\end{equation}
where, as discussed above, this should be understood as the probability that an event labeled $\bx_B$ by Bob is labeled $\bx_A$ by Alice.

As an example, on the semiclassical gravitational state $|\eta\ket$ representing Minkowski space, appropriate initial conditions for the observers lead to a \emph{Lorentz boost} along $x$,
\begin{align}
t(\bx_\B;\eta) &= \gamma(t_\B - \beta x_\B) \, , \\
x(\bx_\B;\eta) &= \gamma(x_\B - \beta t_\B) , \\
y (\bx_\B;\eta)  & = y_\B , \\
z (\bx_\B;\eta)  & = z_\B\, .
\end{align}
In the next section we will discuss how the above is modified by the presence of an impulsive gravitational wave.

Now suppose we consider a \emph{superposition} of  semiclassical states $|g\ket$. In general we do not expect them to form an orthonormal basis. The point is that in  gravity the spatial metric $h$ is the canonical variable. Its eigenstates are not well-behaved space\emph{times} because  they have maximal uncertainty on the momentum $\pi \sim \dot h$. In linearized gravity however we have a firm handle on states representing a classical field configuration:  these are  the coherent states of the graviton. So we expect $|g\ket$ to be some kind of overcomplete coherent state basis in the general case. 
Useful discussions of these states are found in~\cite{Berezhiani:2020pbv,Berezhiani:2021zst}, and in~\cite{Papadodimas:2015jra} with AdS/CFT insights.

The inner product $\bra g|g'\ket$ makes an interesting topic of its own, beyond the scope of this paper. In the linear theory this quantity is  calculable, as we show in App.~\ref{coherent}. For plane waves with a common wavefront  it is exponentially suppressed by the transverse area of the waves in Planck units, see eq.~\eqref{inner}.  Similarly, in~\cite{Papadodimas:2015jra}, it is argued that semiclassical metrics are orthogonal in asymptotically AdS up to  corrections of order $\sim e^{-  N}$, where $N$ scales as $\sim \ell^2/G$ and $\ell$ is the AdS radius. 
In what follows we calculate~\eqref{probability} for generic values of $\bra g|g'\ket$ and then consider the limit where $\bra g|g'\ket \sim \delta(g - g')$. 
This greatly simplifies our formulas but still gives non-trivial results, as we will see in the next subsection. Finally, we should note that this orthogonal limit is obtained in the case of JT gravity where the gravitational degrees of freedom are encoded in a quantum variable defined on the boundary of the theory. The eigenvalues of this variables $|f\ket$ already represent a smooth spacetime which is a portion of $AdS_2$.% We consider this system in Sec.~\ref{JT}. 

If only a single semiclassical state $\bar{g}$ leads to a specific coordinate transformation  $\bx_\B \rightarrow \bx_\A$, then  the projector $Q_{\bx_B\rightarrow \bx_A}$ appearing in (\ref{ct3})  can be written as: 
\be \label{ct4}
Q_{\bx_B\rightarrow \bx_A} = |\bar{g}\ket \bra\bar{g}|. 
\ee
We generally expect that more than one semiclassical metric $g$ can produce some given transformation $\bx_\B \rightarrow \bx_\A$ and that these metrics form a subspace. Then, in the semiclassical limit in which the coherent states are roughly orthogonal,   the  projector will have the general form:
\begin{equation} \label{proj}
Q_{\bx_B\rightarrow \bx_A} = \int d g \,  \delta^4(\bx_\A - \bx(\bx_\B;g))\  n(\bx_\A, g) |g\ket\bra g| \, .
\end{equation}
The measure $dg$ is the generalization of the standard coherent states complex measure $\frac{d^2 \alpha}{\pi}$ for a single harmonic oscillator. It is normalized in such a way that 
\begin{equation} \label{unit}
\mathbb{1} = \int dg |g\ket\bra g|\, .
\end{equation}
In equation  (\ref{proj}),  $n(x_\A,g)$ is the  density of states around $| g\ket$, appropriately normalized so that   the resulting operator is indeed a projector.  The formula for the probabilistic coordinate transformation (\ref{ct3})  says that the probability to realize a coordinate transformation $\bx_\B\rightarrow \bx_\A$ equals that of being on a generic semiclassical metric which realizes such a coordinate transformation.

In practice, it is hard to obtain something more explicit than (\ref{proj}), or otherwise get some handle on the density of states appearing  in that formula,  without further simplifying assumptions. In  the next section we will present a situation  where the probability gets contributions from a single state (at least in a certain ``minisuperspace'' approximation), and can be computed explicitly.  
\subsection{Boosted observers through a gravitational waves superposition }\label{boost}
 As an explicit example we now consider a superposition of impulsive gravitational waves with a common wavefront, which depend on two real parameters $a_+$ and $a_\times$ representing the amplitudes of the two polarizations of the wave. Such classical metrics will be  given explicitly below. 
 The set of semiclassical states we consider are the coherent states corresponding to these metrics. More specifically, we limit ourselves to the states in the subspace they span
 \begin{equation} \label{gen-quantum}
|\psi\rangle = \sum_{h = +,\times}\int da_h \, \psi(a_h) |a_h\rangle\, ,
\end{equation}
This limitation can be thought of as a ``minisuperspace'' approximation. 
 
 In standard ``Brinkmann" coordinates the classical metric reads (see e.g.~\cite{Ferrari:1988cc,Steinbauer:1997dw,Zhang:2017geq,Zhang:2017jma} and references therein)
\begin{equation}
ds^2 = - 2 du dV + f(X,Y) \delta(u) du^2 + dX^2 + dY^2.
\end{equation}
This is a solution of the Einstein equations in the vacuum if 
\begin{equation}
f(X,Y) = a_+ \, (X^2 - Y^2) + 2 a_\times  \, X Y\, .
\end{equation}
In App.~\ref{app_impulse} we review the geodesics of this metric. Of particular relevance are those geodesics representing an initially static observer with Minkowski coordinates $\vec x_\A$ at $u<0$. After the passage of the wave \emph{Alice}-observers receive a ``kick" in Brinkmann coordinates~\cite{Zhang:2017jma}. This means that in the presence of a \emph{superposition} of waves $|\psi\rangle$ the Brinkmann coordinates of the observers end up being undetermined. However, as advertised already, we can \emph{use the $A$-observers to define an event}.  This corresponds to using ``Rosen coordinates" adapted to the $A$-observers. In the $A$-frame, $A$-observers are perfectly localized by definition, and the metric reads, to linear order, 
\begin{align} \label{secondline}
ds^2 =  -& \eta_{\mu \nu} dx_\A^\mu d x_\A^\nu + 2 u \theta(u) \left[ a_+  \left({dx_\A}^2 -   {dy_\A}^2 \right)+ 2 a_\times dx_\A dy_\A\right]\ ,
\end{align}
where $u =(t_\A - z_\A)/\sqrt{2}$.
We can study geodesics in these coordinates. We consider a free falling observer with initial velocity $v$ along the $x_\A$ direction.  
Before hitting the wave (at $u<0$), their trajectories read
\begin{equation}
x_\A(t_\A) = \frac{x_\B}{\gamma} - v \, t_\A, \quad y_A(t_A) = y_\B, \quad z_\A(t_\A) = z_\B , 
\end{equation}
where $\gamma$ is the relativistic gamma factor and  $\vec x_\B$ label the generic boosted observer. In order to get something more similar to a Lorentz transformation for $u<0$, we need to express these trajectories as a function of the observers' proper time $t_\B$, and synchronize that time such that $x_A(t_B=0)=x_B$. The full geodesic in terms of this proper time then reads\footnote{As these are first order formulas, they should not be trusted at distances from the wave larger than $\sim a_{+\times}^{-1}$. 
The fact that free falling particles do not appear to follow straight lines at $u>0$ is because we are using ``distorted coordinates". The spacetime is clearly Minkowski at $u>0$ and all geodesics are straight lines when expressed in Brinkmann coordinates (see App.~\ref{app_impulse}).}
\begin{align}
t(\bx_\B;a_h) &= \gamma(t_\B - v x_\B) - \theta(u) a_+  \sqrt{2}v^2\  u^2/2\, \label{t} \\[1mm]
x(\bx_\B;a_h) &= \gamma(x_\B - v t_\B) + \theta(u) a_+ \sqrt{2}  v\ u^2 , \label{x} \\[1mm]
y (\bx_\B;a_h)  & = y_\B + \theta(u) a_\times \sqrt{2}  v \ u^2  , \label{y}\\[1mm]
z (\bx_\B;a_h)  & = z_\B - \theta(u) a_+  \sqrt{2}v^2\ u^2/2 \, , \label{z}
\end{align}
where $u =  (\gamma(t_\B - v x_\B)-z_\B)/\sqrt{2}$.

As an obvious check notice that for $v =0$ we get $\bx = \bx_\B$, regardless of the state of the wave. In general however the positions of the \emph{Bob}-observers in  the \emph{Alice}-frame depend on the two parameters $a_+$ and $a_\times$. As a result, on a gravitational state~\eqref{gen-quantum}, such a position is not certain, but instead associated to a \emph{probability density} $p(\bx_\A|\bx_\B)$ over the $d$-dimensional space of $\bx_\A$ (given a fixed $\bx_\B$).

We now suppose the gravitational field is  in a  pure  state \eqref{gen-quantum},  belonging to the Hilbert subspace spanned by $|a_+\ket$ and $|a_\times\ket$. % In the formula~\eqref{probability-general} this reduces the functional integration over $g_1$ and $g_2$ to a simple integral over the real variables $a_+$ and $a_\times$. 
Then, the probability distribution in equation (\ref{ct3}) is the expectation value of the appropriate projection operator: 
\be \label{prob1}
p(\bx_\A | \bx_\B) = \bra \psi |  Q_{\bx_\B \to \bx_\A}  | \psi \ket. 
\ee 
In this case the projection operator can be written explicitly:   for a given coordinate transformation, i.e. for a given pair of coordinates $\bx_\A ,  \bx_\B$, among all the semiclassical states  labeled by $a_h \equiv (a_+, a_\times)$  there exist at most  one state such that  $\bx_\A ,  \bx_\B$  are connected by the coordinate transformation (\ref{t}-\ref{z}):  it is the one identified by setting
\begin{equation}  \label{ahbar}
\bar{a}_+ = \sqrt{2}\, \frac{z_\B - z_\A}{v^2 u^2} \qquad \bar{a}_\times = \frac{y_\A - y_\B}{\sqrt{2} v u^2} \, 
\end{equation}
as can be obtained by inverting~\eqref{y} and~\eqref{z}. The relevant subspace is then one-dimensional (if we restrict ourselves to the Hilbert space of coherent states) and the corresponding projector is simply: 
\be \label{proj-ah}
 Q_{\bx_\B \to \bx_\A} = | \bar{a}_h \ket \bra \bar{a}_h |. 
\ee
Using this expression in equation (\ref{prob1})  leads to: 
\be \label{prob2}
p(\bx_\A | \bx_\B) = |\bra \psi | \bar{a}_h \ket |^2 = \left|\int d^2a_h' \psi^*(a_h')  \bra a_h' | \bar{a}_h \ket \right|^2. 
\ee
 
In App.~\ref{coherent} we evaluate the scalar product 
\begin{equation}\label{inner}
\langle a_h'|a_h\rangle \ \sim \ \delta_{h h'} \ e^{- (a'_h - a_h)^2 M_P^2 S_T  \Delta l ^2}\, .
\end{equation}
where $S_T$ and $\epsilon$ are two regulators proportional to the transverse area of the wave and its thickness respectively. These  coherent states can be made arbitrarily orthogonal by pushing on the regulators. Also, in the large-$N$ limit of AdS/CFT it is argued~\cite{Papadodimas:2015jra} that  $\bra g|g'\ket \sim \delta(g - g')$ for any semiclassical pair of states\footnote{%Arguably, this means that those semiclassical states $|g_3\ket$ that have sizable inner products with the states $|a_{+\times}\ket$ within our statistical ensemble are so close to $|a_{+\times}\ket$ that the outcome in terms of observers' trajectories are indistinguishable from those of $|a_{+\times}\ket$.
In the context of our problem, we can take this to mean that any pair of $g$ and $g'$ with non-negligible $\braket{g|g}$ are so similar that the paths observers take through the two spacetimes are essentially indistinguishable.}. In this limit, equation (\ref{prob2}) leads to the probability distribution {\em in the space of coherent states:} 
\be \label{prob3}
dP_{\bx_\B \to \bx_\A} = |\psi(\bar{a}_h)|^2 d^2 \bar{a}_h. 
\ee

To connect with our general discussion, we should express the  probability as a distribution over the 4-dimensional space of Alice-frame coordinates $d^4\bx_\A$. This can be done by changing variables from  $a_h$ to $y_A,z_A$ using (\ref{ahbar}) for fixed $\bx_\B $ and inserting  a two-dimensional $\delta$-function to reconstruct the full 4-dimensional measure. This leads to the final result for the probabilistic coordinate transformation:
\be \label{resultP}
dP_{\bx_\B \to \bx_\A} = \delta(t_A -\bar{t}_A)\delta(x_A -\bar{x}_A) {|\psi(\bar{a}_h)|^2  \over v^3 u^4} \, d^4\bx_\A
\ee
where  $\bar{t}_A$ and $\bar{x}_A$ are the functions of  $\bx_\B$ and $a_h$ appearing on the right hand side of (\ref{t}), (\ref{x}), with the $\bar{a}_h$ themselves replaced by their expressions in (\ref{ahbar}). This way, the right hand side of (\ref{resultP}) is a distribution localized on a 2-dimensional surface in 4-dimensional $\bx_\A$-space, and is a function of $\bx_\B $.

Eq.~\eqref{resultP} should be compared with what is expected from a classical deterministic coordinate transformation, where the probability distribution is concentrated at a single point,  
\be \label{pclass}
p_{classical}= \delta^4(\bx_\A - x(\bx_\B,a_h)).
\ee 
Quantum metric fluctuations cause the probability distribution in~\eqref{resultP}  to spread over a 2-dimensional surface\footnote{Here,  the distribution remains localized in two dimensions   because of the restricted GW superposition that we consider~\eqref{gen-quantum}, which has only two independent parameters.  General metric fluctuations are expected to produce probabilistic coordinate transformations which spread in all directions.}.

\section{Dressing} \label{dressing}
Since coordinate transformations become probabilistic in the presence of metric fluctuations it is natural to ask whether something similar happens to local field operators. 
Consider a standard field theory on fixed background that contains a set of local observables ${\cal O}(\bx)$. 
 Under a coordinate transformation $\bx  \rightarrow  {\bx'}(\bx) $ these operators transform  as  
 \begin{align} \label{transf}
{\cal O}(\bx) &\ \longrightarrow \ {\cal O}'(\bx') =  {\cal O}(\bx(\bx'))\, .
\end{align}
Once dynamical gravity is turned on, ${\cal O}(\bx)$ are no longer observables because they fail to be gauge invariant under coordinate transformations, as shown in the equation above. In order to get rid of any ambiguity one can then prescribe a complete gauge fixing and make calculations in that coordinate system. This is daily practice in cosmological perturbation theory.\footnote{It might be worth here seeing what are the effects of the relativity of the event in cosmological perturbation theory. Consider two gauges, each that implies complete gauge fixing (e.g. spatially flat and longitudinal). The coordinates are related by 
\begin{equation}
\bx ' = \bx + {\boldsymbol \xi (\bx)}\, .
\end{equation}
A scalar field (the inflaton) transforms like in~\eqref{transf},  $\phi'(\bx')  = \phi(\bx) - \xi^\mu \partial_\mu \phi(\bx) + \dots$\, .  
The important thing here is that ${\boldsymbol \xi (\bx)}$ is a function of the metric and/or of the scalar field. The reason is that we are considering a transformation between two gauges that are completely determined. 
As a result, correlators of $\phi$, when expressed in some other gauge, are polluted by higher order correlators taken in the coincident limit. E.g.
\begin{equation}
\langle \phi(\bx') \phi(\by')\rangle = \langle \phi(\bx) \phi(\by)\rangle -  \langle \xi^\mu(\bx) \partial_\mu \phi(\bx) \phi(\by)\rangle -  \langle \phi(\bx)  \xi^\mu(\by) \partial_\mu \phi(\by) \rangle + \dots\, .
\end{equation}
We thank Mehrdad Mirbabayi for discussions on this point.  }  Equivalently, one can~\emph{dress} the above operators in order to make them gauge invariant, i.e. combine them with the metric field so that the transformation of the metric field counteracts the transformation~\eqref{transf}.

In general, dressing local operator (as much as complete gauge fixing) can be subtle.  Following ~\cite{Donnelly:2015hta,Donnelly:2016rvo,Giddings:2018umg,Giddings:2019wmj} (see also~\cite{Goeller:2022rsx} for a related approach), we illustrate the basic mechanism by working perturbatively in the gravitational coupling $\kappa = \sqrt{8 \pi G_N}$ around Minkowski space. The graviton field $h_{\mu \nu}$ is normalized canonically, $g_{\mu \nu} = \eta_{\mu \nu} + \kappa h_{\mu \nu}$. Timelike geodesics can be written as perturbations of straight Minkowski trajectories, 
\begin{equation} 
{\boldsymbol x}(s) = s \cdot {\boldsymbol n}+ {\boldsymbol v}(s)  \, .
\end{equation}
In the above $s$ is the proper time, ${\boldsymbol n}$ a future pointing velocity vector and ${\bv}= {\cal O}(\kappa)$. The linearized geodesic equation for ${\boldsymbol v}$ reads
\begin{equation}
n^\rho n^\sigma \left[\partial_\rho \partial_\sigma v^\mu + \frac12 \eta^{\mu \nu}\left(2 \partial_\rho h_{\nu \sigma} - \partial_\nu h_{\sigma \rho}\right)\right] = 0\, .
\end{equation} 
If we work at first order in $\kappa$ it is sufficient to integrate this equation backward in time along the straight line ${\boldsymbol x}(s) = s \cdot {\boldsymbol n}$. We obtain a generalization of the expressions found  in~\cite{Donnelly:2015hta} to arbitrary directions ${\boldsymbol n}$, 
\begin{equation} \label{v}
v^\mu_{{\boldsymbol n}}({\boldsymbol x}) = - \kappa \, \eta^{\mu \nu} \left(\int_0^{\cal A} ds \, \, n^\rho h_{\nu \rho}({\boldsymbol x} - s \, {\boldsymbol n}) + \frac12 \int_0^{\cal A} ds' \int_{s'}^{\cal A} ds \, \, n^\rho n^\sigma h_{\rho \sigma, \nu}({\boldsymbol x} - s \, {\boldsymbol n})\right) + {\cal O}(\kappa^2)\, .
\end{equation}
The integrals in~\eqref{v} extend up to some past anchor surface ${\cal A}$ where, by assumption, the fluctuations of the metric vanish. This is obviously a strong assumption, which seems to require a ``spacelike holographic boundary" in the past of our spacetime. In the \emph{ad hoc} example of  Sec.~\ref{boost} the spacetime is Minkowski except for metric fluctuations propagating at the speed of light from the past null boundary. In a finite volume region of spacetime it is then possible to define the vector $\bv$ by simply propagating the geodesics backward in time and anchoring them anywhere in the initially Minkowski patch. A more robust anchoring is the (suitably renormalized) boundary of asymptotically AdS spacetime, from which we are free to fire our dressing geodesics in the timelike direction---in this case, they rapidly approach null rays in the bulk. This example is discussed in Sec.~\ref{JT} in the context of JT gravity.

Equipped with $\bv$, we can dress the operators by defining
 \begin{equation} \label{cirro}
 \hat{\cal O}_\bn (\bx) = {\cal O}(\bx+\bv_\bn) \simeq \, {\cal O}(\bx) + v^\mu \partial_\mu {\cal O}(\bx) + \dots ,
 \end{equation}
where $ \hat{\cal O}_\bn (\bx)$ is the dressed version of  ${\cal O}(\bx)$. This is a \emph{timelike dressing} in the notation of~\cite{Giddings:2019wmj}. One can check that, under an infinitesimal coordinate transformation, the change  of $\bv$ compensates that of the scalar~\eqref{transf}. In practice, what we are doing is just evaluating ${\cal O}$ 
at a prescribed  position along the geodesic.

Obviously, different timelike dressings are possible, as the combination of the metric entering in $ \hat{\cal O}_\bn (\bx)$ depends on the velocity vector $\bn$. However, the relation between, say,  $ \hat{\cal O}_{\bn_A} (\bx)$ and $ \hat{\cal O}_{{\bn}_B} (\bx)$ is not particularly illuminating. If we probe the geometry of spacetime through the aid of some test field $\phi$, then it could make sense to partially trace over gravity and obtain operators that only act on the Hilbert space of $\phi$.
 If the system happens to be in a product state $|\Psi \ket = |\Psi_h \ket \otimes |\Psi_\phi\ket$ (in the Heisenberg picture this could be a statement about the initial asymptotic state at $t\rightarrow -\infty$)  one can define partially averaged operators 
\begin{equation} \label{partialaverage}
\overline {\cal O}_\bn(\bx) = \bra \Psi_h| \hat {\cal O}_\bn(\bx) |\Psi_h\ket\, ,
\end{equation}
that can also eventually be averaged on $|\Psi_\phi\ket$ to obtain the expectation value $\langle \hat {\cal O}_\bn(\bx) \rangle$.
Formally we can write~\eqref{cirro} as a sum of direct product operators 
 \begin{equation} 
 \hat{\cal O}_\bn(\bx) = \int d^4 y \, \delta^4(\bf x + \bv_\bn (\bx) - \by) {\cal O}(y)  ,
 \end{equation}
so that the partial average in~\eqref{partialaverage} only hits the deltas. We thus get
 \begin{equation} 
 \overline{\cal O}_\bn(\bx) = \int d^4 y \, f_\bn(\bx, \by) {\cal O}(y)  ,
 \end{equation}
where 
\begin{equation}
f_\bn(\bx, \by) \equiv \bra \Psi_h| \delta^4(\bx + \bv_\bn ({\bx }) - {\by} ) | \Psi_h \ket\, .
\end{equation}

If $f(\bx, \by)$ is invertible we can directly relate averaged dressed operators that belong to two different dressing prescriptions, say, $\bn_A$ and $\bn_B$. By defining the kernel 
\begin{equation}
f_{A\rightarrow B}(\bx,\by) = \int d^4 z\ f_B(\by, \bz)\, f^{-1}_A(\bz, \bx),  
\end{equation}
we obtain
\begin{equation} \label{transdress}
\overline{\cal O}_B(\bx) = \int d^4 y \, f_{A\rightarrow B}(\bx,\by) \, \overline{\cal O}_A(\by)\, . 
\end{equation}

The non-local convolution~\eqref{transdress} is the improved version of the point-wise transformation~\eqref{transf} in the presence of metric fluctuations. When changing frame, the $\overline{\cal O}$ operators transform by mixing up with other nearby operators, within the support of the function $f_{A\rightarrow B}$. Eq.~\eqref{transdress} parallels, in spirit, the probabilistic coordinate transformations in the previous section, with a single point-event getting ``spread out" when moving from one frame to another. 
Unfortunately, the ``bar" operators to which~\eqref{transdress} applies are of limited use. The fact that they are state-dependent is somewhat expected in quantum gravity~\cite{Papadodimas:2015jra}. However, as mentioned, they can be defined only if the system happens to be in a product state. Moreover, notice that in order to calculate correlators of dressed operators one has to take the partial average over the entire set of 
multilocal operators entering the correlators. Schematically,
\begin{equation}
\bra  \hat {\cal O}(\bx_1) \hat {\cal O}(\bx_2)\dots  \hat {\cal O}(\bx_n)\ket = \bra \Psi_\phi| \overline{ {\cal O}(\bx_1) {\cal O}(\bx_2)\dots   {\cal O}(\bx_n)}|\Psi_\phi\ket\, , 
\end{equation}
and, in general, 
\begin{equation}
\overline{ {\cal O}(\bx_1) {\cal O}(\bx_2)\dots   {\cal O}(\bx_n)} \ \neq \ \overline{\cal O}(\bx_1) \overline{\cal O}(\bx_2)\dots   \overline{\cal O}(\bx_n)\, .
\end{equation}
While moving from a frame to another, $A\rightarrow B$, these multilocal partially averaged operators transform with covolutions that are a generalization of~\eqref{transdress}.

\section{Discussion} \label{sec:discussion}

With this paper we aimed at spelling out one particular subtlety that arises in quantum gravity whenever one fixes a coordinate system and attempts a quasi-local ``bulk" formulation of the dynamics.  Metric fluctuations make events frame dependent, not just in the standard sense that they are assigned different coordinates in different coordinate systems, but in the deeper sense that changing frame implies a different definition of \emph{what an event is}.

The example of the black hole horizon discussed in Sec.~\ref{JT} is worth reviewing. %The horizon is defined by propagating light rays backward in time from a region of the boundary identified as \emph{future timelike infinity}. In the two-dimensional case of JT gravity, this boundary region is just the point at $t\rightarrow \infty$ (see Fig.~\ref{fig-2}). 
In JT gravity, it is possible to build a frame where the black hole horizon has  a definite position and appears as a sharp codimension-one surface, for example the BMV localization. In those frames built from in-falling observers launched from the boundary, however, the horizon looks smeared. 
The smearing is due to the fact that, at a given proper time of some observer, having reached the horizon yet is an occurrence that happens with some probability, rather than a yes/no statement.
We calculated the variance associated to such a probability distribution to quadratic order in the boundary theory's quantum parameter in Section \ref{JT}.

Boundary calculations of the entropy have gone a long way towards a resolution of the information paradox (see~\cite{Almheiri:2020cfm} and references therein). It is possible that the mechanism that is described here could help us understand the paradox from the local point of view of the observers living in the spacetime and performing experiments while falling inside the black hole. The considerations that apply to the black hole horizon apply also to any geometric object ``made of points", which must be defined relative to some frame and will generally appear smeared in other frames. In particular, the way we define \emph{regions of space} is frame-dependent. This should perhaps be taken into account when trying to understand how the Hilbert space factorizes (e.g.~\cite{Piazza:2005wm,Giddings:2015lla,Donnelly:2016auv,Witten:2023qsv}) in the presence of dynamical gravity.

Causality is another issue impacted by the present discussion. Commonly, it is loosely stated that commutators of local operators develop non-vanishing tails outside the lightcone when gravity is dynamical. In fact, this statement is frame dependent and can be made much more precise. Let us first consider two-dimensional models, such as JT gravity, to begin with. The BMV frame preserves the standard causal structure of classical spacetime---commutators evaluated in that frame \emph{strictly vanish} outside the lightcone~\cite{Blommaert:2019hjr}. Roughly, the reason is that the BMV frame is built/defined by taking the intersections of ingoing and outgoing light rays.

More precisely, one local characterization of causality is given by the lightcone calculated in different frames. By looking at the line element in the BMV frame~\eqref{metricBMV} one sees immediately that $ds^2 = 0$ implies $\Delta t_1 = 0$ or $\Delta t_2 = 0$. On the other hand, in the spacelike localization, from~\eqref{spacelikemetric} we find that the lightcone is given locally by the relation
\begin{equation}
d\bar L = -\left(\frac{\ell\,  \ddot T_b}{\dot T_b} \pm e^{- \bar L/\ell}\right) dt\, .
\end{equation}

Because of its dependence on $T_b$, the first term in parenthesis is fluctuating and the slope of the lightcone $\frac{d \bar L}{dt}$ has a non-zero variance in this frame.

In higher dimensions we do not expect to find a frame like the BMV frame where classical causality is strictly preserved because there are not enough null coordinates to entirely define an event.  The actual value of the commutators and the strength of the apparent causality violations will depend on the frame where the local operators are defined.

Note also that we must be careful of the thought experiment we have in mind when discussing causality. Consider two observers in the JT model leaving the boundary at different times, the earlier observer sending two photons to the later observer. If the second photon is sent quickly enough after the first, then event smearing will cause the light cones to overlap. Naively, this implies that the photons can be received in reverse time order of how they are sent. This, however, neglects the fact that the smearing of the lightcones is correlated. For example, if the first photon is emitted from a location slightly deeper into the bulk than is expected, then so will the second photon. In this thought experiment, the emission events are entangled, so the order is preserved. We might, on the other hand, split the sending of the photon between two experiments. In the first experiment, the first observer sends one photon at one propertime, and in the second experiment, they send another photon at a slightly later propertime. In this case, the second observer in the second experiment could receive the photon at an earlier propertime than they did in the first experiment. Indeed, this anomalous causality can never be observed with just one experiment in general, since any test of the causal structure collapses the spacetime (or in this case, boundary particle) to a particular state\footnote{We thank the referee for raising this interesting point.}.

It is worth noting that most of our calculations in JT gravity have been limited to quadratic perturbations around the black hole saddlepoint. Using the tools developed in~\cite{Harlow:2021dfp}, however, it might be possible to quantify some of the discussed effects completely non-perturbatively. For example, one could compute the commutators between differently dressed operators, quantifying in some loose sense their mutual immeasurability. One could also probe the lightcone structure of a localization using the commutator between operators which are equivalently dressed but which act at different locations in the spacetime\footnote{We thank D. Harlow for discussions on these points.}.

We can also relate the discussion in the paper  with the phenomenon of \emph{space-time nonadditivity} defined in~\cite{Piazza:2022amf}. In a given frame one can calculate the root mean square distance $ \bar d(x,y) =\sqrt{ \langle d^2(x,y)\rangle}$ between any two points $x$ and $y$. We expect that the surface $ \bar d(x,y) = 0$ at a given $x$ can roughly encode the emergent causal relations \emph{in that frame}. Of course, from a fluctuating metric we expect to derive a fluctuating distance. However, anomalous properties show up already at the level of the average distance $\bar d$, because this quantity fails to be the geodesic distance of any classical manifold. This anomaly can be encoded in the \emph{spacetime nonadditivity function} $\mathcal{C}(x;y)$, which is strictly zero when calculated for a standard geodesic distance. In the presence of dynamical gravity $\mathcal{C}$ generally starts at fourth order in a coordinate expansion~\cite{Piazza:2022amf}. Roughly speaking, $\mathcal{C}$ measures the failure of $\bar d$ to be additive along a geodesic. Depending on how one defines a frame, $\mathcal{C}$ can vanish by construction along some directions. If one uses a set of geodesic observers to define the coordinates, for example, there is no anomaly along the time direction because we have used the (additive) observer clocks to define time. So in that case we have~\cite{Piazza:2022amf} $\mathcal{C}(\vec x, t_1;\vec x, t_2) = 0$.

In JT gravity we have calculated $\mathcal{C}$ for various frames. Its complicated expression is not particularly illuminating but it confirms that, depending on the frame/localization chosen, nonadditivity vanishes along certain directions. For example in the BMV localization $\mathcal{C} \propto \Delta t_1^2 \Delta t_2^2$ and thus strictly vanishes along events that are at the same $t_1$ or $t_2$. This is another way to see that the BMV localization preserves the classical causal structure, which is generally lost in other frames. 
We leave a further study of non-additivity in different localizations for future work.

When  building states in quantum gravity, one would like to go beyond  the \emph{ad-hoc}  states considered in Sec.~\ref{prob}. In this work, we only used linearized solutions, i.e. the state is the superposition of different coherent states each one associated to a macroscopic gravitational wave. One can ask whether there exist setups where one can concretely apply  our approach in Sec.~\ref{prob} to states which are the quantum superposition of different semiclassical saddle points which differ at the full nonlinear level. In the context of holography, it may be possible to do such a computation on the field theory side in cases where the true ground state of the dual QFT does not correspond to a single classical saddle of the gravitational theory, but to a superposition of two or more classical solutions. An example of this situation  has been argued to arise in the context of holographic QFTs defined on Euclidean spheres \cite{Ghosh:2021lua}: it may happen that there are two classical saddles in the bulk (the AdS vacuum and a Coleman-de Luccia instanton) which respect the same asymptotic boundary conditions, and are thefore states in the same QFT. Due to the finite volume of space, the true ground state is a superposition of the two. It would be interesting to reproduce this phenomenon in the context of JT gravity (coupled to extra scalars), where one may hope to do an explicit calcuation of the true ground state and obtain   probabilistic coordinate transformations beyond the linearized approximation.

\bigskip
\noindent{\textbf{Acknowledgments:}} We thank Daniel Harlow, Philipp Hoehn, Thomas Mertens, Mehrdad Mirbabayi, Pietro Pellecchia, Andrew Tolley and Edward Witten for useful discussions or correspondence and especially Andreas Blommaert for feedback on an earlier version of this draft.  The work of FP and AT received support from the French government under the France 2030 investment plan, as part of the Initiative d'Excellence d'Aix-Marseille Universit\'e - A*MIDEX (AMX-19-IET-012). The work of FP and AT  is also supported by the Programme National GRAM of CNRS/INSU with INP and IN2P3 co-funded by CNES.

\appendix
\addtocontents{toc}{\protect\setcounter{tocdepth}{1}}

\section{Coherent States} \label{coherent}
Consider the (canonically normalized) graviton field operator\footnote{Here follows some notation and conventions. 
\begin{align}
(dp) & \equiv \sum_{h=+,\times} \frac{d^3p}{(2 \pi)^3 2 E_p}, \quad [\alpha_h(p), \alpha_{h '}^\dagger(p')] = (2\pi)^3 (2  E_p) \delta^3(\vec p- \vec p\, ') \delta_{h h'}, \\ 
\epsilon^+_{ij}(p^3) & =  \begin{pmatrix}
1&0&0\\
0&-1&0\\
0&0&0
\end{pmatrix}, \quad 
\epsilon^\times_{ij}(p^3) =  \begin{pmatrix}
0&1&0\\
1&0&0\\
0&0&0
\end{pmatrix}.
\end{align}
}
\begin{equation}
\hat h_{ij}(x) = \int (dp) \ \epsilon^h_{ij}(\vec p)\left[\alpha_h(\vec p) e^{i  p \cdot x} + \alpha^\dagger_h(\vec p) e^{-i  p \cdot  x}\right],
\end{equation}
Replacing the operators $\alpha_h(\vec p)$ with the scalar function $z_h(\vec p)$ turns the above into a classical configuration. To each classical configuration $z_h(\vec p)$ we can associate a state in the free theory
\begin{align}
    | z_h(\vec p) \ket = \exp \left(-\frac{1}{2}\int (dp) |z_h(\vec p)|^2  \right) \exp \left(\int (dp) z_h(\vec p) \alpha_h(\vec p) \right) | 0 \ket\, ,
\end{align}
which one can show (using the commutation relations) is an eigenstate of the annihilation operators
\begin{align}
    \alpha_{h'}(\vec q) |z_h(\vec p) \ket = z_{h'}(\vec q) |z_h(\vec p) \ket \, .
\end{align}
This property implies
\begin{align}\label{generalCoherentStateInnerProduct}
    \braket{z'_h(\vec p)|z_h(\vec p)} = \exp  \int (dp) \left( -\frac{1}{2} |z_h(p)|^2   -\frac{1}{2} |z'_h(p)|^2 + z'_h(p)^* z_h(p)\right)\, .
\end{align}

A remarkable property of coherent states is that, given any functional $F$ of $\alpha$ and $\alpha^\dagger$, 
\begin{equation} \label{property}
\bra z_h(\vec p)| F[\alpha,\alpha^\dagger]|z_h(\vec p)\ket = \bra 0 | F[\alpha + z,\alpha^\dagger+ z^*]|0\ket\, .
\end{equation}
This makes coherent states the closest analogue to a ``classical" state in the free quantum theory.
Consider, for example, the trajectory of a particle in the presence of a linearized gravitational field
\begin{equation} 
{\boldsymbol x}(s) = s \cdot {\boldsymbol n}+ {\boldsymbol v}(s)  \, .
\end{equation}
where $\bv$ is an operator linear in $\hat h_{ij}$ given in eq.~\eqref{v}. According to the above, the expectation value of ${\boldsymbol x}$ on a coherent state is just the corresponding classical trajectory, and its uncertainty is minimal, i.e., the same as that calculated on the vacuum. So, traveling on a coherent state of gravitons, the particle follows an almost classical trajectory.

We now turn to the particular example from section \ref{boost}. To linear order in the amplitudes $a_h$, the classical metric configuration~\eqref{secondline} is
\begin{equation}\label{standardform}
h_{ij}^{cl}(x) = 2 a_h M_P \epsilon_{ij}^h(p_z) \theta(t-z)(t-z)\,,
\end{equation}
where we have multiplied the metric element by $M_P$ in order to canonically normalize the field. 
It is straightforward to calculate the corresponding $z_h(\vec p)$, 
\begin{align} \label{defef}
z_h(\vec p) = \ & a_h M_P f(\vec p)\, \\
\equiv \ & a_h M_P (2 \pi)^2 \delta(p_x) \delta(p_y) \sqrt{2}\left[- 2 \frac{\theta(p_z)}{p_z} + i \pi \left(\delta (p_z) - |p_z| \delta'(p_z)\right)\right]\,.
\end{align}
Coherent states associated to such a configuration are naturally written as $|af\ket$. Consider two such coherent states, but with different amplitudes $a_h$ and $a'_h$. Using ~\eqref{generalCoherentStateInnerProduct}, their overlap is
\begin{equation}
    \braket{a' f|a f} = \exp \left(- \frac{M_P^2}{2}  \left((a_+'-a_+)^2 + (a_\times'-a_\times)^2 \right) \int (dp) |f(p)|^2 \right)
\end{equation}
The integral in the exponent needs to be regularized. It has IR divergences in the transverse, $xy$, directions. By regularizing the squares of the delta functions as usual as $\delta^2(p_x) \sim \delta(p_x) L_x$ with $L_x$ an IR cutoff, the integral in $p_x$ and $p_y$ produces the transverse area of the wave $S_T$. Analogous dimensional considerations show that the integral over $p_z$ gives something proportional to the thickness of the wave squared, $\Delta l^2$.  The unspecified $O(1)$ factors are regularization dependent and will therefore be omitted. We finally obtain
\begin{equation}
 \braket{a' f|a f} \ \sim \ \ e^{- \sum_h (a'_h - a_h)^2 M_P^2 S_T  \Delta l^2}\, ,
\end{equation}
which reduces to~\eqref{inner} for states of definite polarization. 
Notice that the result is real by construction (the two states share the same spacetime dependence and only differ by their amplitudes) and becomes a delta function in the classical limit.

\section{Impulsive gravitational waves and their geodesics} \label{app_impulse}

We now consider a metric of the form
\begin{equation}
ds^2 = - 2 du dV + f(X,Y) \delta(u) du^2 + dX^2 + dY^2,
\end{equation}
which is a solution of the full vacuum Einstein equations if $\delta^{IJ} \partial_I \partial_J f = 0$, with $I = X,Y$. Clearly, this describes a spacetime which is flat everywhere except along $u=0$. We thus consider a wave with generic polarization
\begin{equation}
f(X,Y) = a_+(X^2 - Y^2) + 2 a_\times X Y\, .
\end{equation}
The non-vanishing Christoffel symbols are
\begin{equation}
\Gamma_{uu}^V = -\frac12 f \, \dot \delta(u), \quad \Gamma^{I}_{uu} = \Gamma^{V}_{I u} = -\frac12 \partial_I f \, \delta(u).
\end{equation}
It is convenient to write the geodesic equation using $u$ as an affine parameter,
\begin{align}
\ddot X &= \frac12 f_X \,  \delta(u)\, ,\\
\ddot Y &= \frac12 f_Y\, \delta(u)\, ,\\
\ddot V &= \frac12 f\, \dot \delta(u)  + (\dot X f_X + \dot Y f_Y)  \, \delta(u) \, . 
\end{align}
While the first two equations can be integrated straightforwardly, the equation for $V$ requires some care in regularizing the delta functions. In particular, one should keep in mind the distributional identities $\dot \delta(u) g(u) = \dot \delta(u) g(0) - \delta(u) \dot g(0)$ and $\theta(u) \delta (u) g(u) = \frac12 \delta(u) f(0)$. Then one gets (e.g.~\cite{Ferrari:1988cc,Steinbauer:1997dw})
\begin{align}
X(u) &=  X_0 + \dot X_0 u + \frac12 f_X(X_0,Y_0) u \theta(u)\, ,\\[1mm]
Y(u) &=  Y_0 + \dot Y_0 u + \frac12 f_Y(X_0,Y_0) u \theta(u)\, ,\\
V(u) &=  V_0 + \dot V_0 u+ \frac12 f(X_0,Y_0) \theta(u) \\
&\ \ + \frac12 \left[ \dot X_0 f_X(X_0,Y_0) + \dot Y_0 f_Y(X_0,Y_0) +\frac14\left( f_X(X_0,Y_0)^2 +  f_Y(X_0,Y_0)^2 \right)\right]u \theta(u) \, .
\end{align}
Notice that $X$ and $Y$ are non differentiable in $u=0$ while $V$ is discontinuous. The above family of geodesics depend on six initial conditions, $X_0$, $\dot X_0$ etc., which we will always choose to set somewhere within the region $u<0$. An easy choice is that of an observer initially at rest starting from the Minkowski coordinate position $(x,y,z)$
\begin{align}\nonumber
X(u) &= x + (a_+ x + a_\times y) u \theta(u)\, ,\\ \label{obsA}
Y(u) &= y + (a_\times x - a_+ y) u \theta(u) \, ,\\
V(u) &= \frac{t+z}{\sqrt{2}}  + \frac12\left[a_+(x^2 - y^2) + 2 a_\times x y\right]  \theta(u) + (a_+^2 + a_\times^2) (x^2 + y^2) u \theta(u) \, . \nonumber
\end{align}
By substituting $u =(t - z)/\sqrt{2}$ in the above we easily obtain the geodesic in terms of the initial coordinate $z$ and proper time $t$ of the observer. By inverting the above equations one can use $x, y, z$ and $t$ ($x^\mu$ in brief) as coordinates for the entire spacetime\footnote{The $x^\mu$ or $x'^\mu$ coordinates are both examples of ``Rosen coordinates" (see e.g.~\cite{Zhang:2017geq}). We have checked that neither class of geodesics develop caustics. This is probably a gift of using \emph{impulsive} GWs, as more general GW solutions are known to encounter coordinate singularities when expressed in Rosen coordinates.}. These coordinates remain attached to the corresponding observer regardless of the polarization and intensity of the wave.

Alternatively, we can consider a new observer starting from the same position, but departing with an initial velocity $v$ along the $x$ direction
\begin{align}\nonumber
X(u) &= \gamma(x' - v t') + \left[a_+ \left(\frac{x'}{\gamma} - v z' \right) + a_\times y'\right] u \, \theta(u)\, ,\\ \label{obsB}
Y(u) &= y' + \left[a_\times \left(\frac{x'}{\gamma} - v z' \right) - a_+ y'\right]u \, \theta(u) \, ,\\
V(u) &=  \frac{\gamma(t' - v x')+z'}{\sqrt{2}}  + \left[\frac{a_+}{2}\left(\left(\frac{x'}{\gamma} - v z' \right)- y'^2\right) +  a_\times \left(\frac{x'}{\gamma} - v z' \right) y' \right]  \theta(u) \nonumber \\
& + \left[-\sqrt{2} v  \left(a_+ \left(\frac{x'}{\gamma} - v z' \right) + a_\times y'\right) + \frac{a_+^2 + a_\times^2}{2} \left(\left(\frac{x'}{\gamma} - v z' \right)^2 + y'^2\right) \right] u \, \theta(u) \, . \nonumber
\end{align}
where one can see the familiar Lorentz boost structure in the region $u<0$. The observer is now labeled by the triplet  $x', y', z'$ , with $u =  (\gamma(t' - v x')-z')/\sqrt{2}$ the relation between the $u$ coordinate and the observers proper time $t'$.

Like before, the primed  coordinates $x', y', z', t'$ (${x'}^\mu$ in brief) follow our boosted observers and cover the entire spacetime.  By inverting~\eqref{obsB} one can relate the primed and unprimed coordinates. 
The resulting expressions are complicated, but can be simplified by expanding to first order in $a_+$ and $a_\times$, 
\begin{align}
x &= \gamma(x' - v t') + \sqrt{2} a_+ v\ \theta(u) u^2\\[1mm]
y &= y' + \sqrt{2} a_\times v \ \theta(u) u^2 \\
z &= z'  - \frac{\sqrt{2}a_+ v^2}{2} \ \theta(u) u^2 \\
t &= \gamma(t' - v x') - \frac{\sqrt{2}a_+ v^2}{2}\  \theta(u) u^2\, ,
\end{align}
%where, again, $u =  (\gamma(t' - v x')-z')/\sqrt{2}$.

\section{Averages and variances}\label{varianceAppendix}
The coordinates of the horizon in some given frame are generally a functional of boundary Poincar\'e time $T_b(t)$. In particular, they appear to be of the form $F(T_b, \dot T_b)$. We will compute variances of such quantities perturbatively at lowest order in $\epsilon$ around the black hole saddlepoint. It is useful to find a general formula that we can then apply repeatedly. In order to shorten the notation here we substitute $T_b(t)\rightarrow q(t)$ and $T_{b,\text{BH}}(t) \rightarrow q_0(t)$ in our formulae.

A straightforward Taylor expansion around the classical solution gives
\begin{align}
  F(T_b,\dot T_b) = F + \frac{\partial F}{\partial q} \delta q + \frac{\partial F}{\partial \dot q} \delta \dot q + \frac{1}{2} \frac{\partial^2 F}{\partial q} \delta q^2 + \frac{1}{2} \frac{\partial^2 F}{\partial \dot q} \delta \dot q^2 + \frac{\partial^2 F}{\partial q \partial \dot q} \delta q \delta \dot q + O(\delta q^3)
\end{align}
The variance of this quantity is
\begin{align}
\Delta F\equiv    \braket{F^2} - \braket{F}^2 = \Big< \Big(\frac{\partial F}{\partial q} \delta q + \frac{\partial F}{\partial \dot q} \delta \dot q \Big)^2 \Big> - \Big< \Big(\frac{\partial F}{\partial q} \delta q + \frac{\partial F}{\partial \dot q} \delta \dot q \Big) \Big>^2 
\end{align}
Since $q(t)= q_0(t+\epsilon(t))$ we have at first order $\delta q = \dot  q_0 \, \epsilon $ and $\delta q = \ddot q_0 \, \epsilon + \dot q_0 \, \dot \epsilon $. We get
\begin{align} \label{simpto}
   \Delta F = \Big< \Big(\dot F \epsilon + \frac{\partial F}{\partial \dot  q} \dot q_0 \dot \epsilon \Big)^2 \Big> -  \Big< \dot F \epsilon + \frac{\partial F}{\partial \dot q} \dot q_0 \dot \epsilon \Big>^2 
\end{align}
On a classical saddlepoint $\braket{\epsilon}=\braket{\dot \epsilon}=0$, and for the saddlepoint considered in the main text  $\braket{\epsilon(t)\dot \epsilon(t)}=0$. This simplifies~\eqref{simpto} to
\begin{align}
      \Delta F = \dot F^2 \braket{\epsilon^2} + \Big( \frac{\partial F }{\partial \dot q } \dot q_0 \Big)^2\braket{\dot \epsilon^2} \, .
\end{align}

\section{Geodesic localizations}\label{appendixPerpLoc}
\subsection{Perpendicular localization}
Here we show that, at next to leading order in $\delta/\ell$, the localization given by \eqref{spacelikeGeo}-\eqref{spacelikeGeo2} is equivalent to the one where the observers leave perpendicular to the boundary. Spacelike geodesics in $AdS_2$ are given by
\begin{align}
    T &= T_0 + \frac{1}{E} \left( \frac{1}{\tanh( - \tau/\ell + A)} - \frac{\ell \dot Z_0}{Z_0} \right)\, ,
    \\
    Z &= \frac{1/E}{\sinh(-\tau/\ell + A)}\, ,
\end{align}
where
\begin{align}
   E^2 = \frac{\ell^2}{Z_0^2} \left(\frac{\dot Z_0^2}{Z_0^2} - \frac{1}{\ell^2} \right) = \frac{\ell^2 \dot T_0^2}{Z_0^4} 
\end{align}
and
\begin{align}
    A = \frac{1}{2} \log \left( \frac{\sqrt{E^2  Z_0^2 + 1} + 1}{\sqrt{E^2  Z_0^2 + 1} - 1} \right) = \frac{1}{2} \log\left(\frac{\ell \dot Z_0 + Z_0}{\ell \dot Z_0 - Z_0} \right)\, ,
\end{align}
with $(T_0,Z_0)$ and $(\dot T_0,\dot Z_0)$ the initial position and velocity in Poincar\'e coordinates. We want to leave perpendicular to the boundary, meaning we should have $(\dot T_0, \dot Z_0) \propto (\dot Z_b , \dot T_b)$. Normalizing
\begin{align}
    (\dot T_0 , \dot Z_0) = \sqrt{\frac{-Z_b^2/\ell^2}{-\dot T_b^2 + \dot Z_b^2}}(\dot T_b, \dot Z_b) = \frac{\delta}{\ell} (\dot Z_b , \dot T_b) + O(\delta^3)
\end{align}
and so
\begin{align}
    \frac{1}{E} &= \frac{\dot T_b^2}{\ddot T_b} + O(\delta^2)\, ,
    \\
    e^{-A} &= \frac{1}{2} E Z_b + O(\delta^3)\, .
\end{align}
It is straightforward to show from here that the geodesics to leading order are
\begin{align}
    T &= T_b + O(\delta^2)\, 
    \\
    Z &= Z_b e^{\tau/\ell} + O(\delta^2) \, .
\end{align}

\subsection{Tangential localization}\label{tangentialLoc}
Here we show that, at next to leading order in $\delta/\ell$, the localization given by \eqref{timelikeGeo1}-\eqref{timelikeGeo2} is equivalent to the one where the observers leave tangential to the boundary. Timelike geodesics in $AdS_2$ are given by
\begin{align}
    T &= T_0 + \frac{1}{E} \left( \tan (\tau/\ell + B) - \frac{\ell \dot Z_0}{Z_0} \right)\, 
    \\
    Z &= \frac{1/E}{\cos(\tau/\ell + B)} \, 
\end{align}
where
\begin{align}
    E^2 = \frac{\ell^2}{Z_0^2} \left(\frac{\dot Z_0^2}{Z_0^2} + \frac{1}{\ell^2} \right) = \frac{\ell^2 \dot T_0^2}{Z_0^4} 
\end{align}
and
\begin{align}
    B = \tan^{-1} \left( \sqrt{E^2 Z_0^2 - 1} \right) = \tan^{-1} \frac{\ell \dot Z_0}{Z_0}\, ,
\end{align}
with $(T_0,Z_0)$ and $(\dot T_0,\dot Z_0)$ the initial position and velocity in Poincar\'e coordinates. We want to leave tangential to the boundary, meaning we should have $(\dot T_0, \dot Z_0) \propto (\dot T_b , \dot Z_b)$. Normalizing
\begin{align}
    (\dot T_0, \dot Z_0) = \sqrt{\frac{-Z_b^2}{-\dot T_b^2 + \dot Z_b^2}}(\dot T_b, \dot Z_b) = \frac{\delta}{\ell}(\dot T_b, \dot Z_b) + O(\delta^3)
\end{align}
and therefore
\begin{align}
    \frac{1}{E} = \delta \dot T_b + O(\delta^3)\, 
    \\
    B = \frac{\delta \ddot T_b}{\dot T_b} + O(\delta^3)\, .
\end{align}
It is again straightforward to show that
\begin{align}
    T &= T_b + \delta \dot T_b \tan(\tau/\ell) + O(\delta^2)\, ,
    \\
    Z &= \frac{\delta \dot T_b}{\cos(\tau/\ell)} + O(\delta^2) \, .
\end{align}

\renewcommand{\baselinestretch}{1}\small
\bibliographystyle{ourbst} %REMOVED DUE TO BIBLATEX
%\bibliography{replicaBib} %REMOVED DUE TO BIBLATEX
\bibliography{references} %REMOVED DUE TO BIBLATEX

%\printbibliography %BIBLATEX ADDITION

\end{document}